\batchmode
\makeatletter
\def\input@path{{C:/Users/rober/Documents/Doctorado/Paper/Entrega/Arxiv/}}
\makeatother
\documentclass[onecolumn,english,british,twocolumn]{IEEEtran}
\usepackage[T1]{fontenc}
\usepackage[latin9]{inputenc}
\usepackage{float}
\usepackage{amsmath}
\usepackage{amsthm}
\usepackage{amssymb}
\usepackage{graphicx}

\makeatletter

\providecommand{\tabularnewline}{\\}
\floatstyle{ruled}
\newfloat{algorithm}{tbp}{loa}
\providecommand{\algorithmname}{Algorithm}
\floatname{algorithm}{\protect\algorithmname}

\theoremstyle{plain}
\newtheorem{thm}{\protect\theoremname}
\theoremstyle{plain}
\newtheorem{lem}[thm]{\protect\lemmaname}
\theoremstyle{definition}
\newtheorem{example}[thm]{\protect\examplename}

\usepackage{cite} 
\usepackage[margin=8pt,font=footnotesize]{caption}
\usepackage{algorithm}
\usepackage{algpseudocode}
\usepackage{amsmath}
\usepackage{tikz} 
\allowdisplaybreaks

\ifdefined\showcaptionsetup
 \PassOptionsToPackage{caption=false}{subfig}
\fi
\usepackage{subfig}
\makeatother

\usepackage{babel}
\addto\captionsbritish{\renewcommand{\examplename}{Example}}
\addto\captionsbritish{\renewcommand{\lemmaname}{Lemma}}
\addto\captionsbritish{\renewcommand{\theoremname}{Theorem}}
\addto\captionsenglish{\renewcommand{\algorithmname}{Algorithm}}
\addto\captionsenglish{\renewcommand{\examplename}{Example}}
\addto\captionsenglish{\renewcommand{\lemmaname}{Lemma}}
\addto\captionsenglish{\renewcommand{\theoremname}{Theorem}}
\providecommand{\examplename}{Example}
\providecommand{\lemmaname}{Lemma}
\providecommand{\theoremname}{Theorem}

\begin{document}
\title{Variable Dimension IMM and GPB2 Filters for Tracking in Multiple Model
Systems}
\author{Roberto Pérez-Pérez and Ángel F. García-Fernández\thanks{R. P\'erez-P\'erez and A. F. Garc\'ia-Fern\'andez are with the Information Processing and Telecommunications Center, ETSI de Telecomunicaci\'on, Universidad Polit\'ecnica de Madrid, 28040 Madrid, Spain (emails: \mbox{roberto.perez.perez@alumnos.upm.es} and \mbox{angel.garcia.fernandez@upm.es}). This work was supported by the Spanish Ministry of Science, Innovation and Universities under the project PID2024-158149OB-C21.}}
\maketitle
\begin{abstract}
This paper presents mathematically principled filters for multiple-model
systems with states of different dimensionality, specifically the
variable-dimension interacting multiple model (VD-IMM) filter and
the variable-dimension generalised pseudo-Bayesian filter of order
2 (VD-GPB2). To do so, we first provide a Bayesian modelling of a
variable dimensional dynamic system, and its measurements. Then, for
variable dimensional linear Gaussian dynamic and measurement models,
the VD-IMM filter is derived by assuming a posterior that has a Gaussian
density for each mode, and then performing a Kullback-Leibler Divergence
(KLD) minimisation after each prediction step to keep the Gaussian
density form for each mode. Subsequently, the Bayesian update step
is performed, keeping a Gaussian density for each mode. The VD-GPB2
filter considers the same model as the VD-IMM filter and also assumes
that the posterior for each mode is Gaussian. In contrast, the VD-GPB2
filter propagates a Gaussian mixture for each mode in the prediction
step. In the update step, the VD-GPB2 filter performs a KLD minimisation
to have a Gaussian density for each mode. Simulation results show
the benefits of the VD-IMM and VD-GPB2 filters compared to previous
alternatives.
\end{abstract}


\section{Introduction}

Multiple model (MM) filtering consists of estimating a state whose
dynamics evolve across distinct motion regimes, based on noisy measurements
\cite{Challa_book11}. This is particularly important in target tracking
scenarios involving abrupt manoeuvrers, mode switching, or heterogeneous
motion behaviours. Such conditions can arise in different applications
such as plane tracking, perception in autonomous vehicles and maritime
awareness \cite{RongLi2005,Jo2012,Yuan2017,Visina2018,Hem2024,Zhou2025}.

MM filtering is typically posed as a Bayesian filtering problem, in
which the state includes a mode and a kinematic state and we aim to
compute the density of the state given all past observations \cite{Challa_book11}.
Here, the dynamic model is composed of a mode transition density (modelling
changes between modes) and a dynamic model for the kinematic state
of each mode. This paper focuses on the MM filtering problem in which
the state for different modes has a different dimensionality (unequal
state dimension). This is a common characteristic of MM filtering
problems since there are motion parameters that are only relevant
in some modes, but not in others \cite{BarShalom2001}. In the following,
we first review the standard MM filtering approaches, in which the
state for each mode is of fixed dimensionality. Then, we review the
approaches in the literature that deal with variable dimensional (VD)
MM filtering, and then we present the contribution.

Arguably, the most widely used MM filter is the Interacting Multiple
Model (IMM) filter \cite{Blom1988}. Based on linear-Gaussian models,
the IMM filter runs mode-matched Gaussian filters for each mode. In
theory, this leads to a posterior of the state given the mode that
is a Gaussian mixture density due to the possible transitions among
the different modes \cite{BarShalom2001}. To speed up computation,
the IMM filter propagates a single Gaussian for each mode by performing
moment matching in the prediction step, resulting in a mixing step.
The IMM filter can also be applied to nonlinear models \cite{BarShalom2001}.
Other IMM variants include the variable structure IMM, which considers
a variable set of models \cite{Li1996}, and the reweighted IMM, which
modifies how the weights are updated to compute the maximum a posteriori
estimate \cite{Johnston2001}. The IMM filter has also been extended
to distributed filtering, for instance in \cite{Acar2021,Li2023}.

The Generalised Pseudo-Bayesian filter of order 1 (GPB1) \cite{Ackerson1970,BarShalom2001}
is another MM filter that assumes a posterior Gaussian distribution
for the state (without conditioning on the mode). Then, there is a
different prediction for each mode that is collapsed into a Gaussian
distribution again by moment matching.

The Generalized Pseudo-Bayesian filter of order 2 (GPB2) \cite{BarShalom2001},
as the IMM filter, propagates a posterior density approximation that
is Gaussian given each mode. However, instead of projecting the Gaussian
mixture in the prediction step, it does so in the update. This implies
an increase of computational burden compared to the IMM filter, and
typically a slight improvement in performance. A variation of the
GPB2 filter that applies a correction factor based on variational
inference was proposed in \cite{Li2019}.

If we directly apply the standard IMM and GPB2 filters in VD-MM filtering,
the mixing step would require the computation of sums of means and
covariance matrices of different sizes, which are not properly defined
mathematical operations. Therefore, the standard mixing approaches
in the VD-MM literature sort out this problem by augmenting the means
and covariance matrices such that all of them have the same size via
different techniques. For instance, one approach is to augment the
lower dimensional mean and covariance matrices with zeros, and apply
the IMM filter mixing step \cite{BarShalom2001}. This results in
a systematic bias in the additional components of higher dimensional
mode \cite{Yuan12}. To address this problem, an unbiased IMM mixing
step that performs the state augmentation in the mode with lowest
dimensionality taking into account the mean and covariance matrix
of the other mode is proposed in \cite{Yuan12}. In \cite{Granstrom15b},
these IMM mixing approaches for unequal state vectors are generalised
by considering an arbitrary distribution to perform the state augmentation.
Another approach is to perform an IMM mixing step for unequal state
vectors in a reduced common subspace and reconstruct full dimension
states through probability weighted interpolation \cite{Zubaca22}.
This approach avoids explicit augmentation and reduces computational
overhead, although it introduces additional tuning parameters. Further
difficulties of the standard IMM filters arise when there are heterogeneous
motion models having different dimensionality with some state variables
being present in some dynamic models but not in others, e.g., some
models consider velocity in Cartesian coordinates while others consider
velocity in polar coordinates. In fact, the previously mentioned IMM
extensions to states of variable dimensions cannot be applied to this
case. To deal with this, the IMM filter in \cite{Na2022}, augments
the states of some modes such that all modes have the same state variables,
except some modes that can have additional variables, and then the
unbiased IMM mixing step is applied \cite{Yuan12}.

A central drawback of all the above-mentioned VD-MM filters is that
they take an algorithm developed for fixed dimensionality Bayesian
filtering, and sort out in different ways the practical problems that
arise when this algorithm is used in variable dimensional setting
(sums of means and covariances of different dimensionalities). Instead,
in this paper, we take a principled mathematical look at VD-MM filtering
by first posing the VD-MM filtering problem in a Bayesian setting.
To do so, we first define a state space that accounts for variable
dimensionality by defining it as the disjoint union of spaces of different
dimensionality. Spaces of this type have been previously been used
to represent the single-object space and the single-trajectory space
in multiple target tracking using random finite sets \cite{Mahler2014,GarciaFernandez2020}.
Once we have defined the state, we define the dynamic and measurement
densities and obtain the associated Bayesian VD-MM filtering recursion.
An important difference with the fixed dimensional MM filters is that
the transition density must depend on the current mode and the previous
mode to be properly defined. 

To derive the associated linear-Gaussian VD-IMM filter, we perform
Kullback-Leibler divergence (KLD) minimisation on the state space
after the prediction step to keep a Gaussian density for each mode.
To derive the associated linear Gaussian VD-GPB2 filter, we perform
the same KLD minimisation but after the update step. Therefore, the
derivations of the proposed VD-MM filters only require a suitable
state, suitable dynamic and measurement models, and KLD minimisations,
resulting in principled VD-MM filters. Contrary to previous approaches,
the proposed VD-MM filters do not involve sums of means and covariance
matrices of different dimensionalities or require external tuning
parameters or considerations. The proposed filters can also handle
heterogeneous motion models of different dimensionality \cite{Na2022}
seamlessly, without external modifications.

In summary, the contributions of this paper are the development of: 
\begin{enumerate}
\item Principled Bayesian formulation of the variable dimension multiple
model filtering problem.
\item Principled VD-IMM filter based on minimising the KLD to fit a Gaussian
for the state density given each mode at each prediction step.
\item Principled VD-GPB2 filter based on minimising the KLD to fit a Gaussian
for the state density given each mode at each update step.
\item Extension of the VD-IMM and VD-GPB2 filters to nonlinear models via
its extended Kalman filter (EKF) implementation \cite{BarShalom2001}.
\end{enumerate}
The rest of the paper is organized as follows. Section\,\ref{sec:Problem-Formulation}
formalizes the VD-MM filtering problem. Section\, \ref{sec:Variable-Dimension-IMM}
presents the proposed VD-IMM filter, and Section\,\ref{sec:Variable-Dimension-GPB2}
presents the proposed VD-GPB2 filter. Section \ref{sec:Discussion:-Fixed-versus}
discusses the differences between fixed dimension and variable dimension
MM filters. Section \ref{sec:Extension-to-Nonlinear} explains the
EKF implementations of the proposed filters for nonlinear models.
Section \ref{sec:Simulations} provides simulation results and a comparative
performance analysis with respect to other variable dimension filters.
Finally, Section \ref{sec:Conclusions} concludes the paper.

\section{Problem Formulation}\label{sec:Problem-Formulation}

In this section, we provide the problem formulation for principled
variable dimension IMM and GPBP2 filters. First, we present the mathematical
model used for VD-MM Bayesian filtering in Subsection \ref{subsec:Models}.
Then, we provide the Bayesian recursion for variable dimension multiple
model filtering in Subsection \ref{subsec:Bayesian-Filtering-Solution}.

\subsection{Models}\label{subsec:Models}

We consider a single target space with multiple modes $\mathbb{X}=\uplus_{r=1}^{m}\left\{ r\right\} \times\mathbb{R}^{n_{r}}$
where the $r$-th mode dimensionality $n_{r}$ may be different from
$n_{r'}$ for $r\neq r'$ and the symbol $\mathbb{\uplus}$ is used
to represent the union of disjoint sets \cite{Mahler2014}. Thus,
the target state is $\left(r,x\right)\in\mathbb{X}$ where the kinematic
state $x\in\mathbb{R}^{n_{r}}$ and the mode $r\in\left\{ 1,...,m\right\} $.
The target state includes relevant information about the target's
kinematics, such as position, velocity, and angular velocity. The
considered variables depend on the target's motion mode, e.g., constant
velocity or coordinated turn \cite{BarShalom2001}. 

A state $\left(r_{k-1},x_{k-1}\right)$ at time step $k-1$ evolves
to another state at time step $k$ with transition density
\begin{align}
p\left(r_{k},x_{k}|r_{k-1},x_{k-1}\right) & =\mu\left(r_{k}|r_{k-1}\right)\nonumber \\
 & \quad\times\pi\left(x_{k}|x_{k-1},r_{k},r_{k-1}\right)\label{eq:joint_mode_state}
\end{align}
where $\mu\left(\cdot|\cdot\right)$ is the transition density for
the mode, and $\pi\left(\cdot|\cdot\right)$ is the transition density
for the states given the modes. Note that $\pi\left(\cdot|\cdot\right)$
should depend on both $r_{k}$ and $r_{k-1}$ as this is required
to define the transition density on space $\mathbb{X}$ and properly
account for a change of dimensionality. At time step zero, the state
has a prior density $f_{0|0}(r_{0},x_{0})=f_{0|0}(r_{0})f_{0|0}\left(x_{0}|r_{0}\right)$.

The measurement state consists of observable variables provided by
the sensor system, e.g., position and velocity. It is defined as $z\in\mathbb{R}^{n_{z}}$,
where $n_{z}$ denotes the dimension of the measurement space. Thus,
the measurement model is described by the conditional density $l(z_{k}|r_{k},x_{k})$
of the measurement $z_{k}$ given the mode $r_{k}$ and the target
state $x_{k}$.

\subsection{Variable Dimension Multiple Mode Bayesian Filtering}\label{subsec:Bayesian-Filtering-Solution}

This subsection presents the Bayesian recursion for variable dimensional
filtering to compute the posterior joint density of the mode and the
state at time step $k$, $\left(f_{k|k}\left(r_{k},x_{k}\right)\right)$,
conditioned on all relevant prior information up to time $k$. The
predicted and posterior densities of the state at time step $k$ given
the measurements up to time step $k'\in\{k-1,k\}$ are expressed as
\begin{align}
f_{k|k'}(r_{k},x_{k}) & =f_{k|k'}(r_{k})f_{k|k'}\left(x_{k}|r_{k}\right).\label{eq:post_k_1}
\end{align}

Then, the relation between predicted density at time $k$ and the
posterior at time $k-1$ is given by the following lemma.
\begin{lem}[Bayesian prediction]
 \label{lem:bayesian_pred}

Given a posterior density at time step $k-1$ of the form (\ref{eq:post_k_1}),
the predicted density at time step $k$ is 
\begin{align}
f_{k|k-1}\left(r_{k}\right) & =\sum_{r_{k-1}=1}^{m}\mu\left(r_{k}|r_{k-1}\right)f_{k-1|k-1}\left(r_{k-1}\right)\label{eq:mode_prediction}
\end{align}
\begin{align}
f_{k|k-1}\left(x_{k}|r_{k}\right) & =\sum_{r_{k-1}=1}^{m}\alpha_{k|k-1}^{\left(r_{k},r_{k-1}\right)}\nonumber \\
 & \quad\times\int\pi\left(x_{k}|x_{k-1},r_{k},r_{k-1}\right)\nonumber \\
 & \quad\times f_{k-1|k-1}\left(x_{k-1}|r_{k-1}\right)dx_{k-1}\label{eq:state_prediction}
\end{align}
where
\begin{align}
\alpha_{k|k-1}^{\left(r_{k},r_{k-1}\right)} & =\frac{\mu\left(r_{k}|r_{k-1}\right)f_{k-1|k-1}\left(r_{k-1}\right)}{\sum_{r_{k-1}=1}^{m}\mu\left(r_{k}|r_{k-1}\right)f_{k-1|k-1}\left(r_{k-1}\right)}.\label{eq:IMM_alpha}
\end{align}
\end{lem}
Lemma 1 is proved in Appendix \ref{sec:appendix_a}. Parameter $\alpha_{k|k-1}^{\left(r_{k},r_{k-1}\right)}$
is referred to as the mixing probability in the literature \cite{Challa_book11}.
\begin{lem}[Bayesian update]
\label{lem:bayesian_upd}

Given a prediction density at time step $k$ of the form (\ref{eq:state_prediction}),
the posterior density at time step $k$ is
\begin{align}
f_{k|k}\left(x_{k}|r_{k}\right) & =\frac{l\left(z_{k}|r_{k},x_{k}\right)f_{k|k-1}\left(x_{k}|r_{k}\right)}{\int l\left(z_{k}|r_{k},x_{k}\right)f_{k|k-1}\left(x_{k}|r_{k}\right)dx_{k}}\label{eq:update_state}
\end{align}
\begin{align}
 & f_{k|k}\left(r_{k}\right)\nonumber \\
 & =\frac{f_{k|k-1}\left(r_{k}\right)\int l\left(z_{k}|r_{k},x_{k}\right)f_{k|k-1}\left(x_{k}|r_{k}\right)dx_{k}}{\sum_{r_{k}=1}^{m}f_{k|k-1}\left(r_{k}\right)\int l\left(z_{k}|r_{k},x_{k}\right)f_{k|k-1}\left(x_{k}|r_{k}\right)dx_{k}}.\label{eq:update_mode}
\end{align}
\end{lem}
Lemma 2 is proved in Appendix \ref{sec:appendix_a}.

The objective of this paper is to provide computationally efficient
approximations of the variable dimension Bayesian filter for linear
Gaussian models using Gaussian approximations. The first filter, the
VD-IMM filter, is presented in Section \ref{sec:Variable-Dimension-IMM},
and the second filter, the VD-GPB2 filter, is presented in Section
\ref{sec:Variable-Dimension-GPB2}.

\section{Variable Dimension IMM Filter}\label{sec:Variable-Dimension-IMM}

In this section, we propose the VD-IMM filter. We first present the
variable dimension linear and Gaussian models in Subsection \ref{subsec:Models}.
Subsequently, Subsections \ref{subsec:IMM-Prediction} and \ref{subsec:IMM-Update}
detail the prediction and update phases, respectively. Subsection
\ref{subsec:Mode-and-State-Estimation} addresses mode and state estimation.

\begin{figure*}
\centering
\includegraphics[scale=0.54]{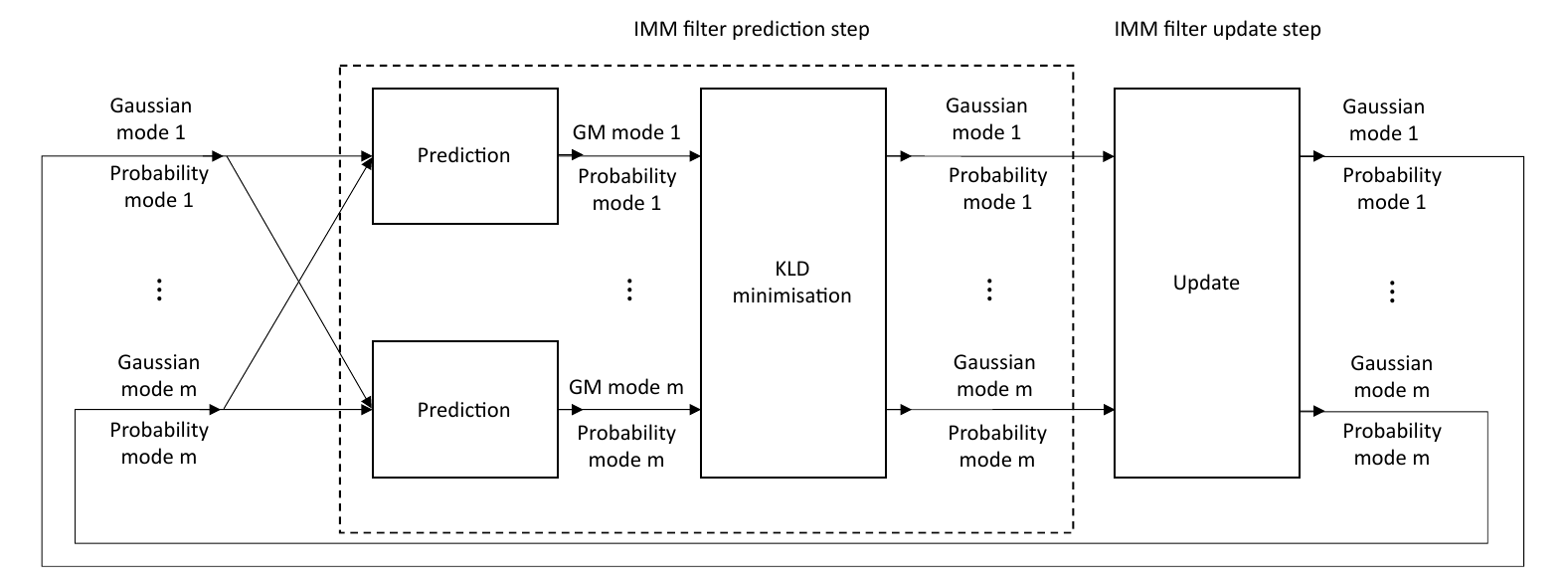}\caption{Diagram of the VD-IMM filter. The prediction step propagates the
mode and the state densities to the current time step. The Bayesian
prediction step results in a Gaussian mixture density for the state
given each mode which is approximated by a Gaussian density by minimising
the KLD on space $\mathbb{X}$. The VD-IMM filter then performs a
Bayesian update, which keeps a Gaussian density for the state given
each mode. The updated density then becomes the input of the prediction
at the next time step.}\label{fig:VD_IMM}
\end{figure*}

\subsection{Dynamic and Measurement Models}\label{subsec:Dynamic-and-Measurement}

Let $\mathcal{N}(x;\bar{x},P)$ denote a Gaussian density with mean
$\bar{x}$ and covariance matrix $P$ evaluated at $x$. In the variable
dimension Gaussian filters, we consider a linear Gaussian transition
model with transition density
\begin{multline}
\pi\left(x_{k}|x_{k-1},r_{k},r_{k-1}\right)\\
=\mathcal{N}\left(x_{k};F^{\left(r_{k},r_{k-1}\right)}x_{k-1}+b^{(r_{k},r_{k-1})},Q^{\left(r_{k},r_{k-1}\right)}\right)\label{eq:transition_model}
\end{multline}
where $F^{\left(r_{k},r_{k-1}\right)}\in\mathbb{R}^{n_{r_{k}}\times n_{r_{k-1}}}$,
$b^{(r_{k},r_{k-1})}\in\mathbb{R}^{n_{r_{k}}}$, $Q^{\left(r_{k},r_{k-1}\right)}\in\mathbb{R}^{n_{r_{k}}\times n_{r_{k}}}$.
Note that these parameters are of suitable dimensions to account for
the state dimension changes.

We consider a linear Gaussian measurement model with likelihood
\begin{align}
l\left(z_{k}|r_{k},x_{k}\right) & =\mathcal{N}\left(z_{k};H^{\left(r_{k}\right)}x_{k}+d^{(r_{k})},R^{\left(r_{k}\right)}\right)\label{eq:measurement_model}
\end{align}
where $H^{\left(r_{k}\right)}\in\mathbb{R}^{n_{z}\times n_{r_{k}}}$,
$d^{(r_{k})}\in\mathbb{R}^{n_{z}}$, $R^{\left(r_{k}\right)}\in\mathbb{R}^{n_{z}\times n_{z}}.$
At time step $0$, the prior $f_{0|0}(\cdot)$ has a Gaussian density
for each mode of the form
\begin{equation}
f_{0|0}\left(x_{0}|r_{0}\right)=\mathcal{N}\left(x_{0};\overline{x}_{0|0}^{\left(r_{0}\right)},P_{0|0}^{\left(r_{0}\right)}\right)\label{eq:time_0_gaussian}
\end{equation}
where $\overline{x}_{0|0}^{\left(r_{0}\right)}$ and $P_{0|0}^{\left(r_{0}\right)}$
are the mean and covariance matrix of $x_{0}$ given $r_{0}$.

\subsection{Prediction}\label{subsec:IMM-Prediction}

In this subsection, we provide the prediction step theorem of the
VD-IMM filter.

Under the linear Gaussian models in (\ref{eq:transition_model}) and
(\ref{eq:measurement_model}), the predicted and updated densities
are of the form (\ref{eq:post_k_1}), with the density of the state
given each mode being a Gaussian mixture. To gain computational efficiency,
the VD-IMM filter assumes that the predicted and updated densities
are of the form (\ref{eq:post_k_1}) with a Gaussian state density
conditioned on the mode of the form
\begin{align}
f_{k|k'}\left(x_{k}|r_{k}\right) & =\mathcal{N}\left(x_{k};\overline{x}_{k|k'}^{\left(r_{k}\right)},P_{k|k'}^{\left(r_{k}\right)}\right)\label{eq:predicted_updated_density}
\end{align}
where $\overline{x}_{k|k'}^{\left(r_{k}\right)}$ and $P_{k|k'}^{\left(r_{k}\right)}$
are the mean and covariance matrix of $x_{k}$ given $r_{k}$ and
measurements up to time step $k'$.

The VD-IMM filter achieves an approximation of the form (\ref{eq:predicted_updated_density})
by minimising the KLD (on space $\mathbb{X}$) between the true predicted
density (with a Gaussian mixture for each mode) and the assumed form
of the predicted density (with a Gaussian for each mode). The expression
of the KLD for densities defined on $\mathbb{X}$ is provided in Appendix
\ref{sec:appendix_b}. A diagram with the VD-IMM filtering recursion
is provided in Figure \ref{fig:VD_IMM}. Then, the VD-IMM prediction
step is given in the following theorem.
\begin{thm}
\label{thm:prediction}(VD-IMM prediction) Given a posterior density
of the form (\ref{eq:post_k_1}) with a Gaussian state density given
each mode (\ref{eq:predicted_updated_density}), the VD-IMM filter
obtains a predicted density of the form (\ref{eq:post_k_1}) with
a Gaussian state density given each mode (\ref{eq:predicted_updated_density})
by minimising the KLD on space $\mathbb{X}$ with parameters
\begin{align}
\overline{x}_{k|k-1}^{\left(r_{k}\right)} & =\sum_{r_{k-1}=1}^{m}\alpha_{k|k-1}^{\left(r_{k},r_{k-1}\right)}\overline{x}_{k|k-1}^{\left(r_{k},r_{k-1}\right)}\label{eq:IMM_x_pred_KLD}\\
P_{k|k-1}^{\left(r_{k}\right)} & =\sum_{r_{k-1}=1}^{m}\alpha_{k|k-1}^{\left(r_{k},r_{k-1}\right)}\nonumber \\
 & \quad\times\left(\overline{x}_{k|k-1}^{\left(r_{k},r_{k-1}\right)}-\overline{x}_{k|k-1}^{\left(r_{k}\right)}\right)\left(\overline{x}_{k|k-1}^{\left(r_{k},r_{k-1}\right)}-\overline{x}_{k|k-1}^{\left(r_{k}\right)}\right)^{T}\nonumber \\
 & \quad+\sum_{r_{k-1}=1}^{m}\alpha_{k|k-1}^{\left(r_{k},r_{k-1}\right)}P_{k|k-1}^{\left(r_{k},r_{k-1}\right)}\label{eq:IMM_P_pred_KLD}
\end{align}
where
\begin{align}
\overline{x}_{k|k-1}^{\left(r_{k},r_{k-1}\right)} & =F^{\left(r_{k},r_{k-1}\right)}\overline{x}_{k-1|k-1}^{\left(r_{k-1}\right)}+b^{(r_{k},r_{k-1})}\label{eq:IMM_x_pred}\\
P_{k|k-1}^{\left(r_{k},r_{k-1}\right)} & =F^{\left(r_{k},r_{k-1}\right)}P_{k-1|k-1}^{\left(r_{k-1}\right)}\left(F^{\left(r_{k},r_{k-1}\right)}\right)^{T}\nonumber \\
 & \quad+Q^{\left(r_{k},r_{k-1}\right)}.\label{eq:IMM_P_pred}
\end{align}

In addition, the predicted density for the mode is given by (\ref{eq:mode_prediction}).
\end{thm}
Theorem \ref{thm:prediction} is proved in Appendix \ref{sec:appendix_c}.
It should be noted that (\ref{eq:IMM_x_pred}) and (\ref{eq:IMM_P_pred})
represent the predicted mean and covariance of the state given a mode
$r_{k}$ at time step $k$ and a mode $r_{k-1}$ at time step $k-1$.

\subsection{Update}\label{subsec:IMM-Update}

In this subsection, we present the update step of the VD-IMM algorithm.
Assuming a Gaussian distribution for the state conditioned on the
mode, see (\ref{eq:predicted_updated_density}), the update step is
closed-form resulting from Bayes' rule.
\begin{thm}
\label{thm:update}(VD-IMM update) Given a predicted density of the
form (\ref{eq:post_k_1}) and (\ref{eq:predicted_updated_density}),
the updated density is also of the form (\ref{eq:predicted_updated_density})
with
\begin{align}
\overline{x}_{k|k}^{\left(r_{k}\right)} & =\overline{x}_{k|k-1}^{\left(r_{k}\right)}+P_{k|k-1}^{\left(r_{k}\right)}\left(H^{\left(r_{k}\right)}\right)^{T}\left(S^{\left(r_{k}\right)}\right)^{-1}\nonumber \\
 & \quad\times\left(z_{k}-\hat{z}^{\left(r_{k}\right)}\right)\label{eq:IMM_x_upd}\\
P_{k|k}^{\left(r_{k}\right)} & =P_{k|k-1}^{\left(r_{k}\right)}-P_{k|k-1}^{\left(r_{k}\right)}\left(H^{\left(r_{k}\right)}\right)^{T}\left(S^{\left(r_{k}\right)}\right)^{-1}\nonumber \\
 & \quad\times H^{\left(r_{k}\right)}P_{k|k-1}^{\left(r_{k}\right)}\label{eq:IMM_P_upd}\\
\hat{z}^{\left(r_{k}\right)} & =H^{\left(r_{k}\right)}\overline{x}_{k|k-1}^{\left(r_{k}\right)}+d^{(r_{k})}\label{eq:IMM_predicted_measurement}\\
S^{\left(r_{k}\right)} & =H^{\left(r_{k}\right)}P_{k|k-1}^{\left(r_{k}\right)}\left(H^{\left(r_{k}\right)}\right)^{T}+R^{\left(r_{k}\right)}.\label{eq:IMM_S_upd}
\end{align}

The posterior of the mode is obtained using (\ref{eq:update_mode}),
which yields
\begin{align}
f_{k|k}\left(r_{k}\right) & =\frac{f_{k|k-1}\left(r_{k}\right)\mathcal{N}\left(z_{k};\hat{z}^{\left(r_{k}\right)},S^{\left(r_{k}\right)}\right)}{\sum_{r_{k}=1}^{m}f_{k|k-1}\left(r_{k}\right)\mathcal{N}\left(z_{k};\hat{z}^{\left(r_{k}\right)},S^{\left(r_{k}\right)}\right)}.\label{eq:update_mode_Gaussian}
\end{align}
\end{thm}
Theorem \ref{thm:update} is proved in Appendix \ref{sec:appendix_c}.
It should be noted that (\ref{eq:IMM_predicted_measurement}) and
(\ref{eq:IMM_S_upd}) represent the predicted measurement and its
covariance matrix conditioned on the state being in mode $r_{k}$.

\subsection{Mode and State Estimation}\label{subsec:Mode-and-State-Estimation}

Mode and state estimation in a VD-IMM filter can be done as follows.
From the updated density, we first estimate the mode with highest
probability
\begin{align}
\hat{r}_{k} & =\underset{r_{k}\in\left\{ 1,...,m\right\} }{\arg\max}f_{k|k}\left(r_{k}\right).\label{eq:mode_estimation}
\end{align}
Then, the state estimate corresponds to the posterior mean conditioned
on $\hat{r}_{k}$, which is $\overline{x}_{k|k}^{\left(\hat{r}_{k}\right)}$. 

Finally, a pseudocode of the VD-IMM filter is provided in Algorithm
\ref{alg:VD-IMM}.

\begin{algorithm}
\textbf{for} $k=1$ \emph{to final step do}
\begin{itemize}
\item Prediction:
\begin{itemize}
\item Calculate $f_{k|k-1}\left(r_{k}\right)$ using (\ref{eq:mode_prediction}).
\begin{itemize}
\item Calculate $\alpha_{k|k-1}^{\left(r_{k},r_{k-1}\right)}$, $\overline{x}_{k|k-1}^{\left(r_{k},r_{k-1}\right)}$
and $P_{k|k}^{\left(r_{k},r_{k-1}\right)}$ using (\ref{eq:IMM_alpha}),
(\ref{eq:IMM_x_pred}) and (\ref{eq:IMM_P_pred}).
\end{itemize}
\item KLD minimisation:
\begin{itemize}
\item Calculate $\overline{x}_{k|k-1}^{\left(r_{k}\right)}$ $P_{k|k-1}^{\left(r_{k}\right)}$
using (\ref{eq:IMM_x_pred_KLD}) and (\ref{eq:IMM_P_pred_KLD}).
\end{itemize}
\item Update:
\begin{itemize}
\item Calculate $\overline{x}_{k|k}^{\left(r_{k}\right)}$, $P_{k|k}^{\left(r_{k}\right)}$,
$\hat{z}^{\left(r_{k}\right)}$ and $S^{\left(r_{k}\right)}$ using
(\ref{eq:IMM_x_upd}) - (\ref{eq:IMM_S_upd}).
\item Calculate $f_{k|k}\left(r_{k}\right)$ using (\ref{eq:update_mode_Gaussian}).
\end{itemize}
\item Estimate the mode $\hat{r}_{k}$ using (\ref{eq:mode_estimation})
and the state $\overline{x}_{k|k}^{\left(\hat{r}_{k}\right)}$.
\end{itemize}
\end{itemize}
\textbf{end for}\caption{Pseudocode of the VD-IMM Filter.}\label{alg:VD-IMM}
\end{algorithm}

\section{Variable Dimension GPB2 Filter}\label{sec:Variable-Dimension-GPB2}

In this section, we present the VD-GPB2 filter. Subsection \ref{subsec:GPB2-Dynamic-and-Measurement}
introduces the required VD dynamic and measurement models. Afterwards,
Subsections \ref{subsec:GPB2-Prediction} and \ref{subsec:GPB2-Update}
develop the prediction and update steps, respectively. Subsection
\ref{subsec:Mode-and-State-GPB2} addresses mode and state estimation.

\begin{figure*}
\centering
\includegraphics[scale=0.54]{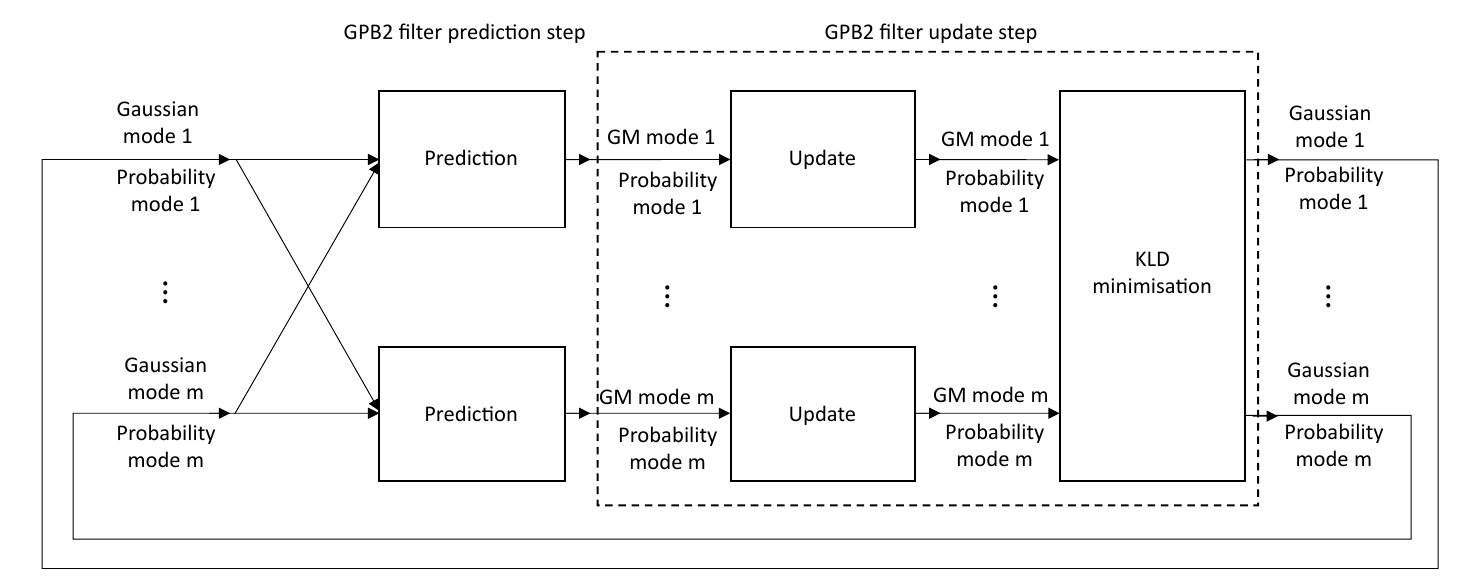}\caption{Diagram of the VD-GPB2 filter. The prediction step propagates the
mode and the state densities to the current time step. The Bayesian
prediction step results in a Gaussian mixture density for the state
given each mode. The output of the Bayesian update step is a Gaussian
mixture density for the state given each mode. The VD-GPB2 filter
then obtains a Gaussian density for the state given each mode by minimising
the KLD on space $\mathbb{X}$. The updated density then becomes the
input of the prediction at the next time step.}\label{fig:VD_GPB2}
\end{figure*}

\subsection{Dynamic and Measurement Models}\label{subsec:GPB2-Dynamic-and-Measurement}

The dynamic and measurement models used in the development of VD-GPB2
are the same as those for the VD-IMM filter, presented in Subsection
\ref{subsec:Dynamic-and-Measurement}.

\subsection{Prediction}\label{subsec:GPB2-Prediction}

In this subsection, we provide the prediction step theorem of the
VD-GPB2 filter.

Assuming the posterior density of the state given the mode at time
step $k-1$ is Gaussian of the form (\ref{eq:predicted_updated_density}),
the predicted density of the state $x_{k}$ given the mode $r_{k}$
is is a Gaussian mixture of the form 
\begin{align}
 & f_{k|k-1}\left(x_{k}|r_{k}\right)\nonumber \\
 & =\sum_{r_{k-1}=1}^{m}\alpha_{k|k-1}^{\left(r_{k},r_{k-1}\right)}\mathcal{N}\left(x_{k};\overline{x}_{k|k-1}^{\left(r_{k},r_{k-1}\right)},P_{k|k-1}^{\left(r_{k},r_{k-1}\right)}\right)\label{eq:predicted_density_gpb2}
\end{align}
where $\alpha_{k|k-1}^{\left(r_{k},r_{k-1}\right)}$, $\overline{x}_{k|k-1}^{\left(r_{k},r_{k-1}\right)}$,
and $P_{k|k-1}^{\left(r_{k},r_{k-1}\right)}$ are given by (\ref{eq:IMM_alpha}),
(\ref{eq:IMM_x_pred}) and (\ref{eq:IMM_P_pred}). The VD-GPB2 filter
performs the Bayesian prediction in Lemma \ref{lem:bayesian_pred}
resulting in the next theorem. 
\begin{thm}
\label{thm:prediction_gpb2}(VD-GPB2 prediction) Given a posterior
density of the form (\ref{eq:post_k_1}) with a Gaussian state density
given each mode (\ref{eq:predicted_updated_density}), the VD-GPB2
filter performs a Bayesian prediction obtaining a density of the form
(\ref{eq:post_k_1}) with a Gaussian mixture state density given each
mode of the form (\ref{eq:predicted_density_gpb2}) with parameters
\begin{align}
\overline{x}_{k|k-1}^{\left(r_{k},r_{k-1}\right)} & =F^{\left(r_{k},r_{k-1}\right)}\overline{x}_{k-1|k-1}^{\left(r_{k-1}\right)}+b^{(r_{k},r_{k-1})}\label{eq:VD-GPB2-state-prediction}\\
P_{k|k-1}^{\left(r_{k},r_{k-1}\right)} & =F^{\left(r_{k},r_{k-1}\right)}P_{k-1|k-1}^{\left(r_{k-1}\right)}\left(F^{\left(r_{k},r_{k-1}\right)}\right)^{T}\nonumber \\
 & \quad+Q^{\left(r_{k},r_{k-1}\right)}\label{eq:VD-GPB2-covariance-prediction}\\
\alpha_{k|k-1}^{\left(r_{k},r_{k-1}\right)} & =\frac{\mu\left(r_{k}|r_{k-1}\right)f_{k-1|k-1}\left(r_{k-1}\right)}{\sum_{r_{k-1}=1}^{m}\mu\left(r_{k}|r_{k-1}\right)f_{k-1|k-1}\left(r_{k-1}\right)}.\label{eq:VD-GPB2-mixing}
\end{align}

In addition, the predicted density for the mode is given by (\ref{eq:mode_prediction}).
\end{thm}
Theorem \ref{thm:prediction_gpb2} is proved in Appendix \ref{sec:appendix_d}.
It should be noted that (\ref{eq:VD-GPB2-state-prediction})-(\ref{eq:VD-GPB2-mixing})
are also computed as part of the VD-IMM filter prediction, see Theorem
\ref{thm:prediction}.

\subsection{Update}\label{subsec:GPB2-Update}

In this subsection, we present the update step of the VD-GPB2 algorithm.
The predicted density of the state for each mode is a Gaussian mixture
of the form (\ref{eq:predicted_density_gpb2}). The Bayesian updated
density for each mode is then another Gaussian mixture, but the GPB2
makes a Gaussian approximation by minimising the KLD on space $\mathbb{X}$.
A diagram of the VD-GPB2 filtering recursion is provided in Figure
\ref{fig:VD_GPB2}.
\begin{thm}
\label{thm:update_gpb2}(VD-GPB2 update) Given a predicted density
of the form (\ref{eq:post_k_1}) with a Gaussian mixture state density
for each mode (\ref{eq:predicted_density_gpb2}), the VD-GPB2 filter
obtains an updated density of the form (\ref{eq:post_k_1}) with a
Gaussian state density for each mode (\ref{eq:predicted_updated_density})
by minimising the KLD on space $\mathbb{X}$ with parameters
\begin{align}
\overline{x}_{k|k}^{\left(r_{k}\right)} & =\sum_{r_{k-1}=1}^{m}\rho^{(r_{k},r_{k-1})}\overline{x}_{k|k}^{\left(r_{k},r_{k-1}\right)}\label{eq:GPB2_x_upd_KLD}\\
P_{k|k}^{\left(r_{k}\right)} & =\sum_{r_{k-1}=1}^{m}\rho^{(r_{k},r_{k-1})}\nonumber \\
 & \quad\times\left(\overline{x}_{k|k}^{\left(r_{k},r_{k-1}\right)}-\overline{x}_{k|k}^{\left(r_{k}\right)}\right)\left(\overline{x}_{k|k}^{\left(r_{k},r_{k-1}\right)}-\overline{x}_{k|k}^{\left(r_{k}\right)}\right)^{T}\nonumber \\
 & \quad+\sum_{r_{k-1}=1}^{m}\rho^{(r_{k},r_{k-1})}P_{k|k}^{\left(r_{k},r_{k-1}\right)}\label{eq:GPB2_P_upd_KLD}
\end{align}
where
\begin{align}
\overline{x}_{k|k}^{\left(r_{k},r_{k-1}\right)} & =\overline{x}_{k|k-1}^{\left(r_{k},r_{k-1}\right)}\nonumber \\
 & \quad+P_{k|k-1}^{\left(r_{k},r_{k-1}\right)}\left(H^{\left(r_{k}\right)}\right)^{T}\left(S^{\left(r_{k},r_{k-1}\right)}\right)^{-1}\nonumber \\
 & \quad\times\left(z_{k}-\hat{z}^{\left(r_{k},r_{k-1}\right)}\right)\label{eq:GPB2_x_upd}\\
P_{k|k}^{\left(r_{k},r_{k-1}\right)} & =P_{k|k-1}^{\left(r_{k},r_{k-1}\right)}\nonumber \\
 & \quad-P_{k|k-1}^{\left(r_{k},r_{k-1}\right)}\left(H^{\left(r_{k}\right)}\right)^{T}\left(S^{\left(r_{k},r_{k-1}\right)}\right)^{-1}\nonumber \\
 & \quad\times H^{\left(r_{k}\right)}P_{k|k-1}^{\left(r_{k},r_{k-1}\right)}\label{eq:GPB2_P_upd}\\
\hat{z}^{\left(r_{k},r_{k-1}\right)} & =H^{\left(r_{k}\right)}\overline{x}_{k|k-1}^{\left(r_{k},r_{k-1}\right)}+d^{(r_{k})}\label{eq:GPB2_z_upd}\\
S^{\left(r_{k},r_{k-1}\right)} & =H^{\left(r_{k}\right)}P_{k|k-1}^{\left(r_{k},r_{k-1}\right)}\left(H^{\left(r_{k}\right)}\right)^{T}+R^{\left(r_{k}\right)}\label{eq:GPB2_S_upd}\\
\rho^{\left(r_{k},r_{k-1}\right)} & \propto\mathcal{N}\left(z_{k};\hat{z}^{\left(r_{k},r_{k-1}\right)},S^{\left(r_{k},r_{k-1}\right)}\right)\nonumber \\
 & \quad\times\mu\left(r_{k}|r_{k-1}\right)f_{k-1|k-1}\left(r_{k-1}\right)\label{eq:rho}
\end{align}
where the proportionality is with respect to $r_{k-1}$ such that
$\sum_{r_{k-1}=1}^{m}\rho^{\left(r_{k},r_{k-1}\right)}=1$. 

The posterior of the mode is obtained using (\ref{eq:update_mode}),
which yields
\begin{align}
f_{k|k}\left(r_{k}\right)=\frac{\sum_{r_{k-1}=1}^{m}\rho^{(r_{k},r_{k-1})}}{\sum_{r_{k}=1}^{m}\sum_{r_{k-1}=1}^{m}\rho^{(r_{k},r_{k-1})}}.\label{eq:update_mode_gpb2}
\end{align}
\end{thm}
Theorem \ref{thm:update_gpb2} is proved in Appendix \ref{sec:appendix_d}.
It should be noted that (\ref{eq:GPB2_z_upd}) and (\ref{eq:GPB2_S_upd})
represent the predicted measurement and its covariance matrix conditioned
on the state being in mode $r_{k}$ at time step $k$ and in mode
$r_{k}-1$ at time step $k-1$. 

\subsection{Mode and State Estimation}\label{subsec:Mode-and-State-GPB2}

The mode and state estimation, used in the VD-GPB2 filter, are the
same that was previously presented in VD-IMM filter (Subsection \ref{subsec:Mode-and-State-Estimation}).
A pseudocode of the VD-GPB2 filter is provided in Algorithm \ref{alg:VD-GPB2}.

\begin{algorithm}
\textbf{for} $k=1$ \emph{to final step do}
\begin{itemize}
\item Prediction:
\begin{itemize}
\item Calculate $f_{k|k-1}\left(r_{k}\right)$ using (\ref{eq:mode_prediction}).
\begin{itemize}
\item Calculate $\alpha_{k|k-1}^{\left(r_{k},r_{k-1}\right)}$, $\overline{x}_{k|k-1}^{\left(r_{k},r_{k-1}\right)}$
and $P_{k|k}^{\left(r_{k},r_{k-1}\right)}$ using (\ref{eq:VD-GPB2-state-prediction}),
(\ref{eq:VD-GPB2-covariance-prediction}) and (\ref{eq:VD-GPB2-mixing}).
\end{itemize}
\item Update:
\begin{itemize}
\item Calculate $\overline{x}_{k|k}^{\left(r_{k},r_{k-1}\right)}$, $P_{k|k-1}^{\left(r_{k},r_{k-1}\right)}$,
$\hat{z}^{\left(r_{k},r_{k-1}\right)}$ and $S^{\left(r_{k},r_{k-1}\right)}$
using (\ref{eq:GPB2_x_upd}) - (\ref{eq:GPB2_S_upd}).
\end{itemize}
\item KLD minimisation:
\begin{itemize}
\item Calculate $\overline{x}_{k|k}^{\left(r_{k}\right)}$ $P_{k|k}^{\left(r_{k}\right)}$
using (\ref{eq:GPB2_x_upd_KLD}) and (\ref{eq:GPB2_P_upd_KLD}).
\item Calculate $f_{k|k}\left(r_{k}\right)$ using (\ref{eq:update_mode_gpb2}).
\end{itemize}
\item Estimate the mode $\hat{r}_{k}$ using (\ref{eq:mode_estimation})
and the state $\overline{x}_{k|k}^{\left(\hat{r}_{k}\right)}$.
\end{itemize}
\end{itemize}
\textbf{end for}

\caption{Pseudocode of the VD-GPB2 filter.}\label{alg:VD-GPB2}
\end{algorithm}

\section{Discussion: Fixed versus Variable Dimension Filters}\label{sec:Discussion:-Fixed-versus}

In this section, we discuss the differences between the standard IMM
and GPB2 methods and their corresponding variable dimensional counterparts.
The first difference between the standard IMM-GPB2 filters and the
VD-IMM-GPB2 filters is that VD-IMM-GPB2 filters consider a state of
variable dimensionality and the associated integrals and densities
from first principles, see Section \ref{sec:Problem-Formulation}
and Appendix \ref{sec:appendix_a}. 

In addition, for principled probabilistic modelling in variable dimensions,
the transition density must be of the form (\ref{eq:joint_mode_state}).
This implies that the linear/Gaussian description of the transition
density on the variable dimensional space requires the parameters
$F^{\left(r_{k},r_{k-1}\right)}$, $b^{\left(r_{k},r_{k-1}\right)}$
and $Q^{\left(r_{k},r_{k-1}\right)}$. That is, these matrices must
depend on the current mode $r_{k}$ and also on the previous mode
$r_{k-1}$ and be of the proper dimensions to perform the change of
dimensionality, see (\ref{eq:transition_model}). In contrast, in
the standard IMM-GPB2 filters, the transition density (\ref{eq:joint_mode_state})
is simplified as
\begin{align}
\pi\left(x_{k}|x_{k-1},r_{k}\right).\label{eq:simplified_transition_density}
\end{align}

In the linear-Gaussian case, this implies that the dynamic model matrices
only depend on $r_{k}$, such that we have $F^{\left(r_{k}\right)}$,
$b^{\left(r_{k}\right)}$ and $Q^{\left(r_{k}\right)}$. In addition,
typically the offset parameter $b^{\left(r_{k}\right)}$ is not considered,
though this is straightforward to add. That is, standard IMM-GPB2
filters simplify the $m\times m$ transition densities $\pi\left(x_{k}|x_{k-1},r_{k},r_{k-1}\right)$
to only $m$. However, the simplification in (\ref{eq:simplified_transition_density})
cannot be done with states of different dimensionality, unless the
probability of changing modes is zero, which is not the case of interest.
We proceed to illustrate this with the following example.

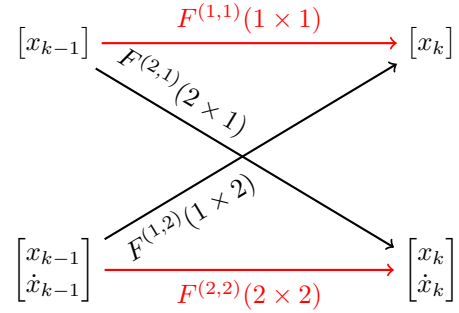
\begin{figure}
\centering

\begin{tikzpicture}[node distance=3cm, auto]
  % Left side nodes (State k-1)
  \node (topleft) {$\begin{bmatrix} x_{k-1} \end{bmatrix}$};
  \node (bottomleft) [below of=topleft] {$\begin{bmatrix} x_{k-1} \\ \dot{x}_{k-1} \end{bmatrix}$};

  % Right side nodes (State k)
  \node (topright) [right of=topleft, xshift=2cm] {$\begin{bmatrix} x_{k} \end{bmatrix}$};
  \node (bottomright) [below of=topright] {$\begin{bmatrix} x_{k} \\ \dot{x}_{k} \end{bmatrix}$};

  % Horizontal arrows (Red)
  \draw [->, red, thick] (topleft) -- (topright) 
    node[midway, above] {$F^{(1,1)}(1 \times 1)$};
  \draw [->, red, thick] (bottomleft) -- (bottomright) 
    node[midway, below] {$F^{(2,2)}(2 \times 2)$};

  % Cross-over arrows (Black)
  % Top-left to bottom-right: label above the line, near the start
  \draw [->, thick] (topleft) -- (bottomright) 
    node[near start, sloped, above] {$F^{(2,1)}(2 \times 1)$};
    
  % Bottom-left to top-right: label below the line, near the start
  \draw [->, thick] (bottomleft) -- (topright) 
    node[near start, sloped, below] {$F^{(1,2)}(1 \times 2)$};
\end{tikzpicture}\caption{State transition diagram (matrix $F^{\left(r_{k},r_{k-1}\right)}$)
for the VD-IMM-GPB2 filters in Example \ref{exa:transition_matrix}.
The red arrows denote transitions between states of fixed dimensionality,
while the black arrows represent transitions between states of different
dimensionality. The different dimensions of matrix $F^{\left(r_{k},r_{k-1}\right)}$
depending on the change of modes are also provided.}\label{fig:State-transition-diagram}
\end{figure}

\begin{example}
\label{exa:transition_matrix}Let us consider $2$ modes, for $r=1,x^{\left(1\right)}=[x]$
(1-D state) and, for $r=2,x^{\left(2\right)}=[x,\dot{x}]^{T}$ (2-D
state). The standard simplification with transition matrices $F^{\left(r_{k}\right)}$
only depending on the mode $r_{k}$ and not $r_{k-1}$ are of size
$1\times1$ and $2\times2$. These constitute an incomplete transition
model when we consider variable dimensional systems since the transitions
between states of different dimensionalities are neglected, see Figure
\ref{fig:State-transition-diagram}. This implies that external procedures
must be used to sort out the resulting inconsistencies in vector dimensionalities
when applying the standard IMM-GPB2 filters. In contrast, directly
using the mathematically principled approach, see (\ref{eq:joint_mode_state})
and (\ref{eq:transition_model}), results in VD-IMM and VD-GBP2 filters
that work seamlessly with different dimensionalities and are shown
to have KLD minimisation properties on the variable-dimensional state,
including mode and kinematic state.
\end{example}
As a result of this modelling difference, in the standard IMM filter,
it is possible to do the moment matching of the Gaussian densities
just before the prediction step for each mode such that we just apply
one prediction with $F^{(r_{k})}$, $Q^{(r_{k})}$ to each mode. A
benefit of this approach is that it simplifies computations since
we do not need to apply $F^{(r_{k},r_{k-1})}$ and $Q^{(r_{k},r_{k-1})}$
to each of the previous modes and the perform moment matching. However,
this simplification does not enable us to do principled VD-MM filtering,
as explained above. Another drawback of having an $F^{(r_{k})}$ only
depending on $r_{k}$ is that even if the state vectors are of the
same dimensionality, but with different representations (e.g., one
model is in Cartesian coordinates, and another is in polar coordinates),
one should have a dynamic model that depends on the current and previous
mode to account for the change in parametrisation. This case can be
handled seamlessly via the VD-MM filters, as demonstrated via simulations
in the next section (Scenario 2). Fixed-dimensional MM filters with
dynamic models depending on $r_{k}$ and $r_{k-1}$ have been previously
considered for instance in \cite{P.Blom1986,RongLi2005}.

In the standard GPB2 filter, the moment matching of the Gaussian densities
is also performed just after the Bayesian update step, as in the VD-GPB2
filter, though the standard GPB2 filter only considers a dynamic model
that depends on $r_{k}$, as explained above.

To address practical target tracking problems, when there is not a
change in mode from time step $k-1$ to $k$, we can choose the dynamic
model for the VD filters similarly to the dynamic model of a fixed
dimensional system, using the relevant physical properties \cite{RongLi2005}.
When there is a change in mode, the dynamic model should be chosen
taken into account the definition of the state in the different modes,
physics and geometry. Two practical examples are provided in the simulation
results in Section \ref{sec:Simulations}.

It should also be noted that if $\mathbb{X}=\uplus_{r=1}^{m}\left\{ r\right\} \times\mathbb{R}^{n}$,
which implies that modes have the same dimensionality ($n$), and
we set $F^{\left(r_{k},r_{k-1}\right)}=F^{\left(r_{k}\right)}$, $b^{(r_{k},r_{k-1})}=0$
and $Q^{\left(r_{k},r_{k-1}\right)}=Q^{\left(r_{k}\right)}$, the
VD-IMM-GPB2 filters reduce to the standard IMM-GPB2 filters.

Finally, it should be mentioned that the standard GPB1 filter \cite{BarShalom2001,Ackerson1970}
uses a Gaussian density to represent the density of the state, across
all modes. This approach is simply not feasible in principled VD-MM
filtering since the densities on the VD space $\mathbb{X}$ cannot
be written in this form.

\section{Extension to Nonlinear Models}\label{sec:Extension-to-Nonlinear}

This section presents the EKF extensions of the VD-IMM and VD-GPB2
filters to nonlinear models. The EKF implementations make a first-order
Taylor series approximation of the nonlinear dynamic and measurement
models \cite{Sarkka_book23} and apply the previously presented VD-IMM
and VD-GPB2 filters for linear models. We first describe the variable
dimensional nonlinear models in Section \ref{subsec:Variable-Dimensional-Nonlinear}.
Then, we describe the EKF VD-IMM filter in Section \ref{subsec:EKF-VD-IMM-Filter}
and the EKF VD-GPB2 filter in Section \ref{subsec:EKF-VD-GPB2-Filter}.

\subsection{Variable Dimensional Nonlinear Models}\label{subsec:Variable-Dimensional-Nonlinear}

We consider a transition density of the form
\begin{multline}
\pi\left(x_{k}|x_{k-1},r_{k},r_{k-1}\right)\\
=\mathcal{N}\left(x_{k};f^{\left(r_{k},r_{k-1}\right)}\left(x_{k-1}\right),Q^{\left(r_{k},r_{k-1}\right)}\right)\label{eq:transition_model_non_linear}
\end{multline}
where $f^{\left(r_{k},r_{k-1}\right)}\left(\cdot\right)$ is the (possibly
nonlinear) dynamic function from mode $r_{k-1}$ to $r_{k}$ that
accounts for dimensionality changes, that is, $f^{\left(r_{k},r_{k-1}\right)}\left(\cdot\right):\mathbb{R}^{n_{r_{k-1}}}\rightarrow\mathbb{R}^{n_{r_{k}}}$
and $Q^{\left(r_{k},r_{k-1}\right)}$ is the process noise covariance
matrix of size $n_{r_{k}}\times n_{r_{k}}$.

The likelihood is of the form
\begin{align}
l\left(z_{k}|r_{k},x_{k}\right) & =\mathcal{N}\left(z_{k};h^{\left(r_{k}\right)}\left(x_{k}\right),R^{\left(r_{k}\right)}\right)\label{eq:measurement_model_non_linear}
\end{align}
where the (possibly) nonlinear measurement function of mode $r_{k}$
is $h^{\left(r_{k}\right)}\left(\cdot\right):\mathbb{R}^{n_{r_{k}}}\rightarrow\mathbb{R}^{n_{z}}$
and $R^{\left(r_{k}\right)}$ is the measurement noise covariance
matrix, of size $n_{z}\times n_{z}$.

For the EKF implementation, we consider that both $f^{\left(r_{k},r_{k-1}\right)}\left(\cdot\right)$
and $h^{\left(r_{k}\right)}\left(\cdot\right)$ are differentiable.

\subsection{EKF VD-IMM Filter}\label{subsec:EKF-VD-IMM-Filter}

Let $F^{\left(r_{k},r_{k-1}\right)}\left(\overline{x}_{k-1|k-1}^{\left(r_{k-1}\right)}\right)$
be the Jacobian matrix \cite{Sarkka_book23} of $f^{\left(r_{k},r_{k-1}\right)}\left(\cdot\right)$
evaluated at a vector $\overline{x}_{k-1|k-1}^{\left(r_{k-1}\right)}\in\mathbb{R}^{n_{r_{k-1}}}$.
Then, the EKF VD-IMM filter makes the following approximation at the
prediction step for all $r_{k}$ and $r_{k-1}$
\begin{align}
F^{\left(r_{k},r_{k-1}\right)} & \approx F^{\left(r_{k},r_{k-1}\right)}\left(\overline{x}_{k-1|k-1}^{\left(r_{k-1}\right)}\right)\label{eq:F_non_linear}\\
b^{\left(r_{k},r_{k-1}\right)} & \approx f^{\left(r_{k},r_{k-1}\right)}\left(\overline{x}_{k-1|k-1}^{\left(r_{k-1}\right)}\right)\nonumber \\
 & \quad-F^{\left(r_{k},r_{k-1}\right)}\left(\overline{x}_{k-1|k-1}^{\left(r_{k-1}\right)}\right)\overline{x}_{k-1|k-1}^{\left(r_{k-1}\right)}.\label{eq:b_non_linear}
\end{align}

The VD-IMM prediction is then performed via Theorem \ref{thm:prediction}
with $F^{\left(r_{k},r_{k-1}\right)}$ and $b^{\left(r_{k},r_{k-1}\right)}$
given by (\ref{eq:F_non_linear}) and (\ref{eq:b_non_linear}).

Let $H^{\left(r_{k}\right)}\left(\overline{x}_{k|k-1}^{\left(r_{k}\right)}\right)$
be the Jacobian matrix \cite{Sarkka_book23} of $h^{\left(r_{k}\right)}\left(\cdot\right)$
evaluated at a vector $\overline{x}_{k|k-1}^{\left(r_{k}\right)}\in\mathbb{R}^{n_{r_{k}}}$.
Then, the EKF VD-IMM makes the following approximation at the update
step for all $r_{k}$ 
\begin{align}
H^{\left(r_{k}\right)} & \approx H^{\left(r_{k}\right)}\left(\overline{x}_{k|k-1}^{\left(r_{k}\right)}\right)\label{eq:H_non_linear_imm}\\
d^{(r_{k})} & \approx h^{\left(r_{k}\right)}\left(\overline{x}_{k|k-1}^{\left(r_{k}\right)}\right)-H^{\left(r_{k}\right)}\left(\overline{x}_{k|k-1}^{\left(r_{k}\right)}\right)\overline{x}_{k|k-1}^{\left(r_{k}\right)}.\label{eq:b_non_linear_imm}
\end{align}

The VD-IMM update is then performed via Theorem \ref{thm:update}
with $H^{\left(r_{k}\right)}$ and $d^{(r_{k})}$ given by (\ref{eq:H_non_linear_imm})
and (\ref{eq:b_non_linear_imm}).

\subsection{EKF VD-GPB2 Filter}\label{subsec:EKF-VD-GPB2-Filter}

The EKF VD-GPB2 filter also performs the approximation (\ref{eq:F_non_linear})
and (\ref{eq:b_non_linear}) in the prediction step and then applies
Theorem \ref{thm:prediction_gpb2}. For the update, let $H^{\left(r_{k}\right)}\left(\overline{x}_{k|k-1}^{\left(r_{k},r_{k-1}\right)}\right)$
be the Jacobian matrix \cite{Sarkka_book23} of $h^{\left(r_{k}\right)}$
evaluated at a vector $\overline{x}_{k|k-1}^{\left(r_{k},r_{k-1}\right)}\in\mathbb{R}^{n_{r_{k}}}$.
Then, the EKF implementation makes this approximation for each $r_{k}$
and $r_{k-1}$
\begin{align}
H^{\left(r_{k}\right)} & \approx H^{\left(r_{k}\right)}\left(\overline{x}_{k|k-1}^{\left(r_{k},r_{k-1}\right)}\right)\label{eq:H_non_linear_gpb2}\\
d^{(r_{k})} & \approx h^{\left(r_{k}\right)}\left(\overline{x}_{k|k-1}^{\left(r_{k},r_{k-1}\right)}\right)-H^{\left(r_{k}\right)}\left(\overline{x}_{k|k-1}^{\left(r_{k},r_{k-1}\right)}\right)\overline{x}_{k|k-1}^{\left(r_{k},r_{k-1}\right)}.\label{eq:d_non_linear_gpb2}
\end{align}
The VD-GPB2 filter update is then performed via Theorem \ref{thm:update_gpb2}
using the approximation (\ref{eq:H_non_linear_gpb2}) and (\ref{eq:d_non_linear_gpb2})
for each $r_{k}$ and $r_{k-1}$. 

\section{Simulations}\label{sec:Simulations}

In this section we evaluate the performance of the VD-IMM and VD-GPB2
filters compared to state-of-the-art IMM filters with states of unequal
size. We consider two scenarios. The first one (Section \ref{subsec:Scenario-1})
corresponds to the standard case in which the states in all modes
have common variables, and the difference in dimensionality is due
to some extra variables in some modes. The second one (Section \ref{subsec:Scenario-2})
corresponds to a case of heterogeneous motion models.

\subsection{Scenario 1}\label{subsec:Scenario-1}

This Subsection evaluates the performance of the proposed VD-IMM (Section
\ref{sec:Variable-Dimension-IMM}) and VD-GPB2 (Section \ref{sec:Variable-Dimension-GPB2})
filters\footnote{Matlab code of the VD-IMM and VD-GPB2 filters is shared via Github
in https://github.com/ropperez/VD\_Multiple\_Models.} against other IMM and GPB2 filters for variable dimensional filtering
in the literature. In Subsections \ref{subsec:Sim_Dynamic-Model},
\ref{subsec:Sim_Mode-Model} and \ref{subsec:Sim_Measurement-Model}
we specify the selected dynamic model, mode model and measurement
model for the carried out simulations. Finally, in Subsections \ref{subsec:Implemented-Algorithms}
and \ref{subsec:Sim_Results} we present the simulation setup, including
all parameter configurations, as well as the results obtained for
each filtering method.

\subsubsection{Dynamic Model}\label{subsec:Sim_Dynamic-Model}

We consider a 2D \emph{coordinated turn} (CT) \cite{BarShalom2001,Challa_book11}
and a nearly constant velocity (CV) \cite{BarShalom2001,Challa_book11}
model pair with unequal state dimensions, where the CT mode includes
turn rate as an additional state. For these simulations, three different
models are selected, CV, coordinated turn to the right (CT\textsubscript{\selectlanguage{english}%
R}) and coordinated turn to the left (CT\textsubscript{\selectlanguage{english}%
L}), which only differ in the offset parameter $b^{\left(r_{k},r_{k-1}\right)}$.
In order to consider changes in the dimensionality of the modes, as
indicated in (\ref{eq:transition_model}), in a nearly constant velocity
model \cite{BarShalom2001} we define the following transition matrices
between modes $F^{(\texttt{{CV,CV}})}$, $F^{(\texttt{{CT,CT}})}$,
$F^{(\texttt{{CV,CT}})}$ and $F^{(\texttt{{CT,CV}})}$. Note that
the superscript CT defines both CT\textsubscript{\selectlanguage{english}%
R} and CT\textsubscript{\selectlanguage{english}%
L} modes for simplicity. 

The state vector for the CV model is defined as $x^{\texttt{{CV}}}=[x,\dot{x},y,\dot{y}]^{T}$
while for the CT models is defined as $x^{\texttt{{CT}}}=[x,\dot{x},y,\dot{y},\omega]^{T}$,
where $x,y$ are the position in Cartesian coordinates, $\dot{x},\dot{y}$
are the lineal velocity in Cartesian coordinates and $\omega$ is
the angular velocity.

For these state vectors, the Markov transition matrices are
\begin{align}
F^{(\texttt{{CV,CV}})} & =I_{2}\otimes F^{(\texttt{{CV}})}\\
F^{(\texttt{{CT,CT}})}\left(x\right) & =\mathrm{diag}\left(F^{(\texttt{{CT}})}\left(x\right),1\right)\label{eq:F_CT_CT}\\
F^{(\texttt{{CT,CV}})} & =\left[\begin{array}{c}
F^{(\texttt{{CV,CV}})}\\
0_{1\times4}
\end{array}\right]\\
F^{(\texttt{{CV,CT}})}\left(x\right) & =\begin{bmatrix}F^{(\texttt{{CT}})}\left(x\right) & 0_{4\times1}\end{bmatrix}
\end{align}
where
\begin{align}
F^{(\texttt{{CV}})} & =\begin{bmatrix}1 & T\\
0 & 1
\end{bmatrix}\\
F^{(\texttt{{CT}})}\left(x\right) & =\begin{bmatrix}1 & \frac{\sin\left(\omega T\right)}{\omega} & 0 & -(\frac{1-\cos\left(\omega T\right)}{\omega})\\
0 & \cos\left(\omega T\right) & 0 & -\sin\left(\omega T\right)\\
0 & \frac{1-\cos\left(\omega T\right)}{\omega} & 1 & \frac{\sin\left(\omega T\right)}{\omega}\\
0 & \sin\left(\omega T\right) & 0 & \cos\left(\omega T\right)
\end{bmatrix}\label{eq:CT_model}
\end{align}
where $T$ is the sampling time interval, $I_{n}$ is the $n\times n$
identity matrix, $0_{m\times n}$ is the $m\times n$ zero matrix,
$\otimes$ is the Kronecker product and $\mathrm{diag}\left(A,B\right)$
creates a block diagonal matrix with blocks $A$ and $B$.

We write the offset parameter, that also depends on the modes, as
$b^{(\texttt{{CV,CV}})}$, $b^{(\texttt{{CV,CT}})}$, $b^{(\texttt{{CT,CT}})}$
and $b^{(\texttt{{CT,CV}})}$. When the current mode is equal to the
previous mode, i.e., $r_{k}=r_{k-1}$, we set
\begin{align}
b^{(\texttt{{CV,CV}})} & =0_{4\times1}\\
b^{(\texttt{{CT,CT}})} & =0_{5\times1}.
\end{align}
When the current mode differs from the previous mode, i.e., $r_{k}\neq r_{k-1}$,
the control parameter is 
\begin{align}
b^{(\texttt{{CV,CT}})} & =0_{4\times1}\\
b^{(\texttt{{CT\_R,CV}})} & =\left[\begin{array}{c}
0_{4\times1}\\
-\omega^{*}
\end{array}\right]\\
b^{(\texttt{{CT\_L,CV}})} & =\left[\begin{array}{c}
0_{4\times1}\\
\omega^{*}
\end{array}\right]\\
b^{(\texttt{{CT\_R,CT\_L}})} & =\left[\begin{array}{c}
0_{4\times1}\\
-\omega^{*}
\end{array}\right]\\
b^{(\texttt{{CT\_L,CT\_R}})} & =\left[\begin{array}{c}
0_{4\times1}\\
\omega^{*}
\end{array}\right]
\end{align}
where $\omega^{*}$ is a known parameter that represents the initial
change in angular velocity when transitioning to a coordinated turn
model. Note that this parameter has a positive sign in the offset
vector when transitioning to a coordinated turn to the left, and a
negative sign when transitioning to a coordinated turn to the right. 

The process noise covariances are \cite{BarShalom2001,Challa_book11}
\begin{align}
Q^{(\texttt{{CV,CV}})}=Q^{(\texttt{{CV,CT}})} & =\sigma_{\texttt{{CV}}}^{2}I_{2}\otimes\begin{bmatrix}\frac{T^{3}}{3} & \frac{T^{2}}{2}\\
\frac{T^{2}}{2} & T
\end{bmatrix}\\
Q^{(\texttt{{CT,CT}})}=Q^{(\texttt{{CT,CV}})} & =\mathrm{diag}\left(Q^{(\texttt{{CV,CV}})},\sigma_{\texttt{{CT}}}^{2}T\right)
\end{align}
where $\sigma_{\texttt{{CV}}}$ and $\sigma_{\texttt{{CT}}}$ are
parameters of the nearly constant velocity model and coordinated turn
model.

It should be noted that when we do not change mode, the transition
matrix is similar to the corresponding fixed-dimensionality models.
However, when there is change from coordinated turn to nearly constant
velocity, $F^{(\texttt{{CV,CT}})}$ changes the size of the state
vector and removes the influence of the angular velocity. In contrast,
when we transition from nearly constant velocity to coordinated turn,
$F^{(\texttt{{CT,CV}})}$ increases the size of the state vector and
the offset vector $b$ provides the initial angular velocity, which
depends on the turn being to the left or to the right. The transition
matrix depending on the state, see (\ref{eq:CT_model}) is handled
using the EKF, as explained in Section \ref{sec:Extension-to-Nonlinear}.

At time step $0$ the mode, mean and covariance of the initial Gaussian
distribution, see (\ref{eq:time_0_gaussian}), are 
\begin{align}
f_{0|0}(\texttt{{CV}}) & =1\\
\overline{x}_{0|0}^{(\texttt{{CV}})} & =\left[0\left(\mathrm{m}\right),50\left(\frac{\mathrm{m}}{\mathrm{s}}\right),0\left(\mathrm{m}\right),50\left(\frac{\mathrm{m}}{\mathrm{s}}\right)\right]^{T}\\
P_{0|0}^{(\texttt{{CV}})} & =\mathrm{diag}\left(\left[1\left(\mathrm{m^{2}}\right),0.1\left(\mathrm{m^{2}}/\mathrm{s}^{2}\right),\right.\right.\nonumber \\
 & \quad\left.\left.1\left(\mathrm{m^{2}}\right),0.1\left(\frac{\mathrm{m^{2}}}{\mathrm{s}^{2}}\right)\right]\right).
\end{align}

\subsubsection{Mode Dynamic Model}\label{subsec:Sim_Mode-Model}

The transition density for the mode, $\mu\left(r_{k}|r_{k-1}\right)$,
introduced in Equation (\ref{eq:joint_mode_state}), is represented
by the following matrix for the variable dimensional filters 
\begin{align}
\begin{bmatrix}0.9 & 0.05 & 0.05\\
0.05 & 0.9 & 0.05\\
0.05 & 0.05 & 0.9
\end{bmatrix}
\end{align}
where the first row and column correspond to the first model (CV),
the second row and column to the second model (CT\textsubscript{\selectlanguage{english}%
R}) and the third row and column correspond to the third model (CT\textsubscript{\selectlanguage{english}%
L}). However, for fixed dimensional filters, the transition density
is defined by the following matrix
\begin{equation}
\begin{bmatrix}0.9 & 0.1\\
0.1 & 0.9
\end{bmatrix}
\end{equation}
where the first row and column correspond to the first model (CV),
the second row and column to the second model (CT\textsubscript{}).

At the beginning of the simulation, we assume that the probability
of the modes is equiprobable, i.e., $f_{0|0}(r_{0})=1/m$, where $m\in\left\{ 2,3\right\} $
denotes the number of the modes. 

\subsubsection{Measurement Model}\label{subsec:Sim_Measurement-Model}

The sensor provides a noisy measurement $z\in\mathbb{R}^{2}$ of the
target position in the $x$ and $y$ coordinates. This leads to the
measurement matrices
\begin{align}
H^{(\texttt{{CV}})} & =\left[\begin{array}{cccc}
1 & 0 & 0 & 0\\
0 & 0 & 1 & 0
\end{array}\right]\\
H^{(\texttt{{CT}})} & =\left[\begin{array}{ccccc}
1 & 0 & 0 & 0 & 0\\
0 & 0 & 1 & 0 & 0
\end{array}\right]
\end{align}
and $d^{\left(r_{k}\right)}=0_{2,1}$. The measurement covariance
matrix $R_{k}$ is 
\begin{equation}
R^{\left(r_{k}\right)}=\sigma_{m}^{2}I_{2}.
\end{equation}

\subsubsection{Parameters}\label{subsec:Implemented-Algorithms}

In this subsection, we provide the values of the parameters introduced
in Subsections \ref{subsec:Sim_Dynamic-Model}, \ref{subsec:Sim_Mode-Model}
and \ref{subsec:Sim_Measurement-Model}, which are used in the simulation
results presented in Subsection \ref{subsec:Sim_Results}.

We provide simulation results with two sampling times, such that sampling
time interval is set to $T\in\left\{ 0.5,1\right\} $ s. The offset
angular velocity is set is fixed to $\omega^{*}=15\times\frac{\pi}{180}$
rad/s. As in \cite{Granstrom15b}, the standard deviation of the process
noise of constant velocity and coordinated turn rate are defined as
\begin{align}
\sigma_{\texttt{{CV}}} & =\gamma\sigma'_{\texttt{{CV}}}\label{eq:sigma_cv}\\
\sigma_{\texttt{{CT}}} & =\gamma\sigma'_{\texttt{{CT}}}\label{eq:sigma_ct}
\end{align}
where $\sigma'_{\texttt{{CV}}}=1$ m/s\textsuperscript{3/2}, $\sigma'_{\texttt{{CT}}}=1\times\frac{\pi}{180}$
rad\textsuperscript{}/s\textsuperscript{1/2} and $\gamma$ is a
dimensionless scaling factor. We will test different values of this
parameters such that
\begin{align}
\gamma & \in\left\{ 10^{-2},10^{-1.75},10^{-1.5},10^{-1.25},10^{-1}\right.\nonumber \\
 & \quad\left.,10^{-0.75},10^{-0.5},10^{-0.25},1,5,10,15,20,25\right\} .
\end{align}
Finally, the standard deviation of the measurement noise intensity
for the Cartesian position coordinates is set to
\begin{equation}
\sigma_{m}\in\left\{ 0.5,1\right\} \text{(m)}.\label{eq:std_measurement_noise}
\end{equation}

\subsubsection{Results}\label{subsec:Sim_Results}

In this subsection, we compare the proposed VD-IMM and VD-GPB2 filters
with the following filters:
\begin{itemize}
\item IMM/GPB2 (standard) \cite{Challa_book11}: these methods augment the
lower dimensional state by assigning zero mean and zero variance to
each missing component, resulting in a biased estimate for the additional
states.
\item IMM-unbiased \cite{Yuan12}: in this strategy, the smaller state vector
is augmented using the mean and variance of the corresponding components
from the higher dimensional state, thereby avoiding the bias introduced
by zero augmentation.
\item IMM-uniform \cite{Granstrom15b}: this method models the missing state
components using a uniform distribution that reflects the expected
range of values for the given scenario. For each augmented state,
an interval $[a,b]$ must be defined to specify the expected value
and variance of the uniform distribution. In this simulation $a=-\omega^{*}$
and $b=\omega^{*}$.
\item IMM-mapping \cite{Zubaca22}: this method introduces a mode mixing
strategy tailored for models whose state vectors have unequal dimension.
The procedure operates in two steps: the common state components are
first mixed, after which the final state estimate is obtained by weighting
the original and mixed states according to the corresponding model
probabilities. This approach has two tunable parameters, $\kappa$
and $\Delta_{0}$, which control the assumed uncertainty of the unmapped
state components and the rate at which the mixing is applied, respectively.
In the simulations, these parameters are set to $\kappa=100$ and
$\Delta_{0}=10$.
\end{itemize}
The results presented in this subsection are obtained through the
averaging of 100 Monte Carlo runs, each consisting of 100 steps per
realisation. The performance comparison among the proposed filters
and previous approaches is carried out in terms of the root mean square
error (RMSE). The RMSE (for positional elements) at time step $k$
as
\begin{align}
\mathrm{{RMSE}_{k}} & =\left(\frac{1}{N_{{MC}}}\sum_{i=1}^{N_{{MC}}}\right.\nonumber \\
 & \quad\left.\times\left[\left(x_{i,k}-\bar{x}_{i,k}^{\left(\hat{r}_{k}\right)}\right)^{2}+\left(y_{i,k}-\bar{y}_{i,k}^{\left(\hat{r}_{k}\right)}\right)^{2}\right]\right)^{\frac{1}{2}}\label{eq:RMSE_k}
\end{align}
while the average RMSE over all realisations and steps is
\begin{align}
\mathrm{RMSE} & =\left(\frac{1}{N}\sum_{k=1}^{N}\mathrm{{RMSE}_{k}}\right)
\end{align}
where $N_{{MC}}$ denotes the number of Monte Carlo runs, $N$ is
the number of time steps of each realisation, $x_{i,k}$ and $y_{i,k}$
represent the ground truth Cartesian coordinates at time step $k$
and Monte Carlo run $i$, $\bar{x}_{i,k}^{\left(\hat{r}_{k}\right)}$
and $\bar{y}_{i,k}^{\left(\hat{r}_{k}\right)}$ are the estimated
positions.

\begin{figure}
\subfloat[\label{fig:ground_truth} Example of two true trajectories generated
for a target with $T=0.5$ s and $\gamma=10^{-2}$ (left) and with
$T=1$ s and $\gamma=10^{-2}$ (right). Blue points correspond to
segments where the dynamic model follows constant velocity, orange
points indicate coordinated turn model to the right, and yellow points
indicate coordinated turn model to the left.]%
{\includegraphics[scale=0.3]{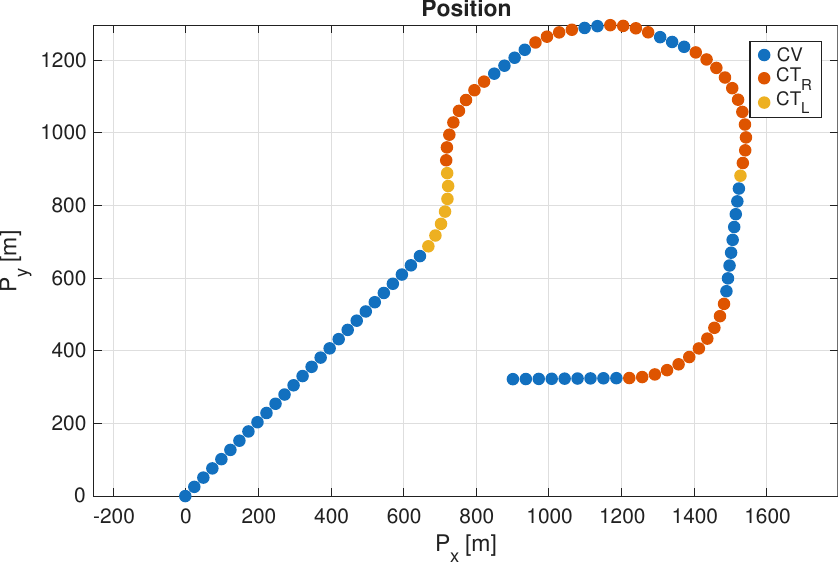}
\includegraphics[scale=0.3]{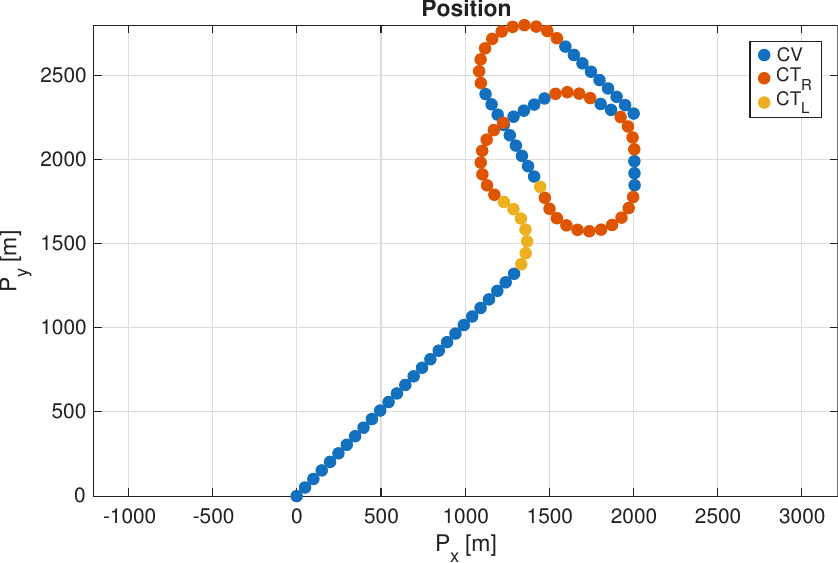}}

\subfloat[\label{fig:Real-trajectory-mode}Real trajectory mode dynamic model.]%
{\includegraphics[scale=0.23]{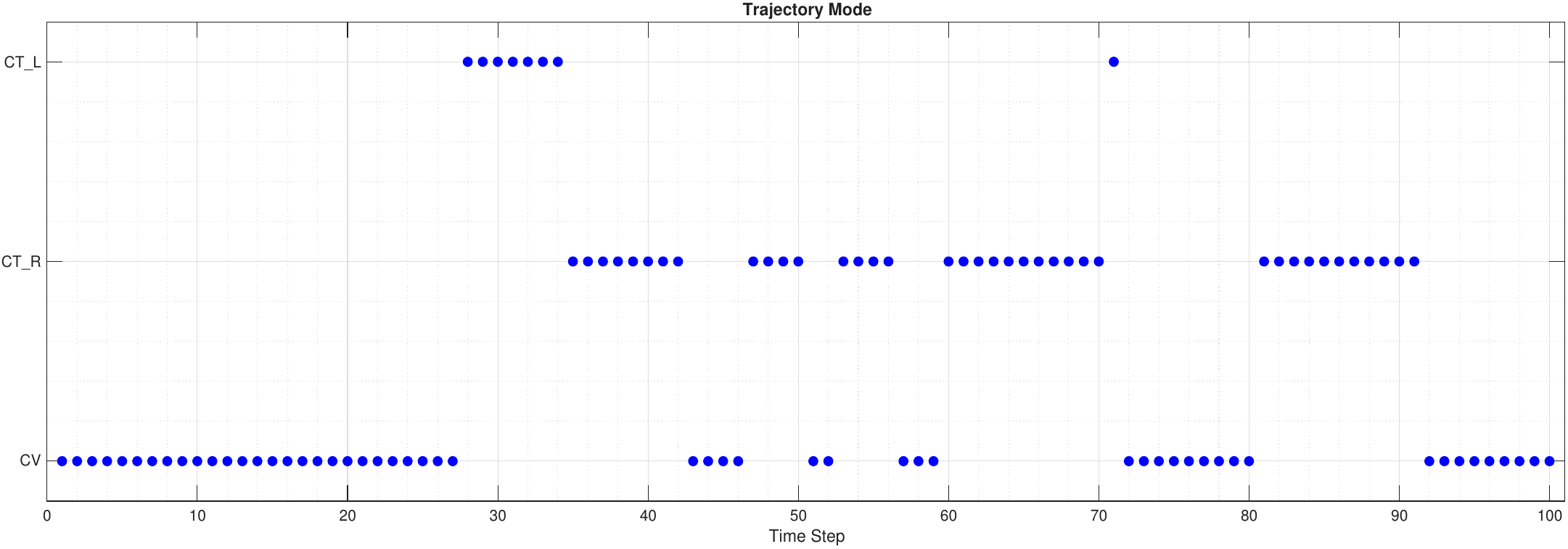}}

\caption{Ground truth of two trajectories with different sampling times (top)
and ground truth mode dynamic model (bottom) for Scenario 1.}
\end{figure}

Each generated trajectory has been drawn from the dynamic model using
the parameters presented above (Subsections \ref{subsec:Sim_Dynamic-Model}
and \ref{subsec:Sim_Mode-Model}), e.g., $T$ and $\gamma$. Consequently,
a different true trajectory is obtained for each parameter set, as
illustrated in Figure \ref{fig:ground_truth}, where two examples
with different parameter configurations are shown, while the bottom
panel shows the ground truth trajectory mode dynamic model as functions
of the time step. It is also important to note that, for all the simulated
filters, the prediction involving the CT mode is implemented using
an EKF \cite{BarShalom2001}.

\begin{figure}
\includegraphics[scale=0.6]{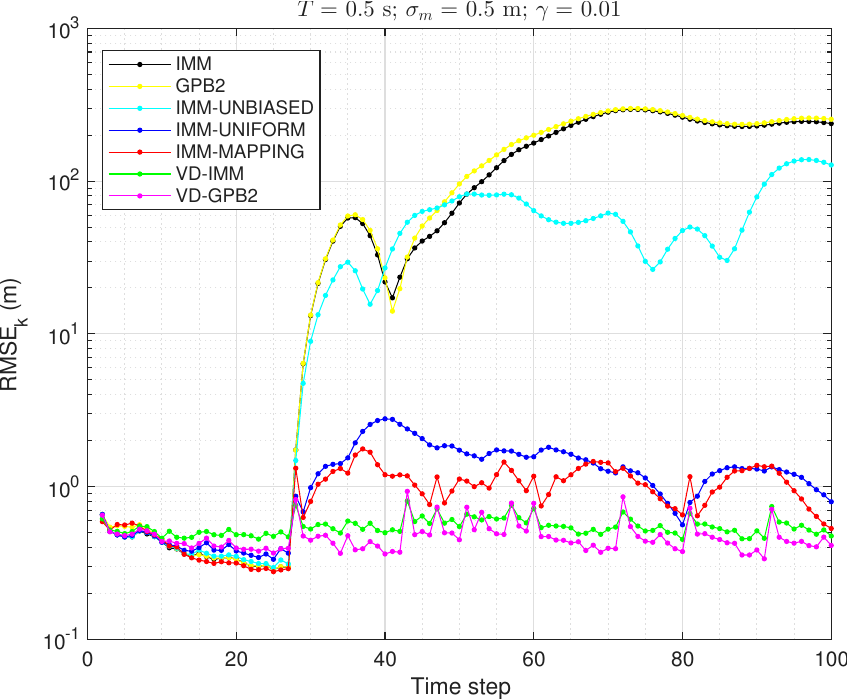}\caption{RMSE for $T=0.5$ s, $\sigma_{m}=0.5$ m and $\gamma=0.01$ for Scenario
1. The VD-GPB2 is the best performing filter followed by the VD-IMM
filter.}\label{fig:RMSE_1_realisation}
\end{figure}

In Figure\,\ref{fig:RMSE_1_realisation}, we show the RMSE against
time obtained for a simulation with a sampling time of $T=0.5$ s,
a process noise level of $\gamma=0.01$, and a measurement noise standard
deviation of $\sigma_{m}=0.5$ m. We can observe that all filters
exhibit an increase in $\mathrm{{RMSE}_{k}}$ whenever a mode transition
occurs, see Figure \ref{fig:Real-trajectory-mode}. A main advantage
of the proposed filters arises from the fact that the simulation is
performed with a non\nobreakdash-zero initial turn rate, $\omega^{*}\neq0$
which can be considered since the dynamic model has information on
the change of mode. If the simulation were carried out with $\omega^{*}=0$,
the benefit of the proposed methods would likely be less pronounced.
From the beginning of the trajectory up to time step $k=27$, all
filters perform almost identically because the mode remains set to
CV. During the transition at time step $k=28$ from CV to CT\textsubscript{\selectlanguage{english}%
R}, the proposed approaches successfully capture the change in the dynamic
model and therefore maintain an almost constant $\mathrm{{RMSE}_{k}}$.
In contrast, the standard IMM, standard GPB2, and IMM-unbiased filters
diverge during this transition, resulting in a significantly higher
$\mathrm{{RMSE}_{k}}$. The IMM\nobreakdash-uniform and IMM\nobreakdash-mapping
filters perform better than these methods; however, they exhibit an
increase in RMSE after the first manoeuvre from which they do not
fully recover. Finally, the proposed VD\nobreakdash-IMM and VD\nobreakdash-GPB2
approaches outperform all other filters, with VD\nobreakdash-GPB2
achieving the best overall performance.

\begin{figure}
\includegraphics[scale=0.6]{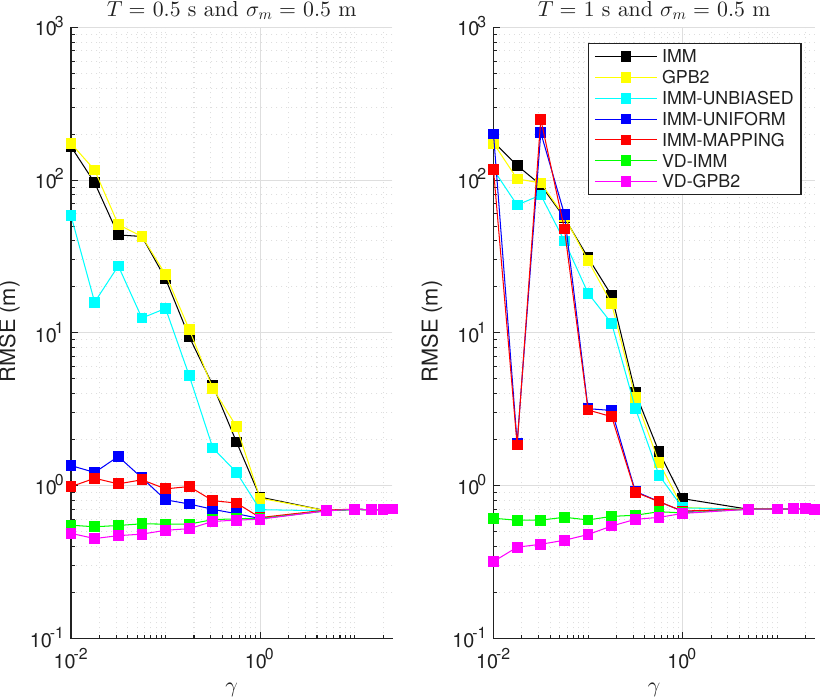}\caption{$\mathrm{RMSE}$ for $\sigma_{m}=0.5$ m and $T=0.5$ s (left) and
with $T=1$ s (right) for Scenario 1. The VD-GPB2 is the best performing
filter followed by the VD-IMM filter.}\label{fig:RMSE_0.5m}
\end{figure}

In Figure \ref{fig:RMSE_0.5m}, we present the results of the average
RMSE across time for different values of $\gamma$ and $T=0.5$ s
and $T=1$ s. We can draw the following conclusions:
\begin{enumerate}
\item When the generated trajectory is dominated by high process noise,
i.e., $\gamma>1$, all filtering methods tend to converge toward similar
RMSE values. In this regime, the estimation error increases for VD
filter variants, while in the other approaches it is reduced. The
reason is that high process noise forces the Kalman gain to rely predominantly
on the measurement rather than on the state prediction. As a consequence,
the filters effectively base their estimation mainly on the measurement
rather than the dynamics. Under these conditions, IMM and GPB2 standard
filters and IMM variants filters exhibit a slight improvement in performance
because their mode dynamic model becomes less relevant when the prediction
step is heavily affected by large process noise. In contrast, the
VD-IMM and VD-GPB2 filters, whose advantage relies on exploiting mode
dynamic model with variable dimension, lose part of their benefit,
leading to a mild degradation in performance relative to the low dynamic
noise scenarios.
\item When the process noise of the generated trajectory is low, the performance
differences among the simulated filters become more pronounced. In
this regime, the Kalman gain must balance the contribution of the
process noise and the measurement noise, forcing the filters to rely
more heavily on the dynamic model. Thus, the ability of each algorithm
to follow the right dynamic model becomes critical. Under these conditions,
the standard IMM and GPB2 filters exhibit the poorest performance,
followed by previous IMM variants. These IMM filter variants do not
fundamentally address the change of dimensionality problem, and cannot
consider the information with the initial angular velocities of the
turns so their improvement is limited. In contrast, the proposed VD-IMM
and VD-GPB2 filters outperform all previous approaches. 
\item When the update interval is increased to $T=1$ s, the RMSE differences
among the methods become higher. The VD-IMM and VD-GPB2 maintain low
RMSE, whereas the standard IMM variants present performance degradation.
A shorter sampling interval provides more frequent updates, lowering
the process noise and therefore improves the overall RMSE.
\item The proposed VD-IMM and VD-GPB2 methods allow the use of separate
filter models for CT\textsubscript{\selectlanguage{english}%
R} and CT\textsubscript{L}, whereas the standard IMM variants rely
on a single CT model for both directions. Therefore, during transitions
from CT\textsubscript{\selectlanguage{english}%
R} to CT\textsubscript{L} or vice versa, the IMM variants struggle
to estimate the change of direction in the turn. This enables the
VD-IMM and VD-GPB2 filters to have better performance.
\end{enumerate}
\begin{figure}
\includegraphics[scale=0.6]{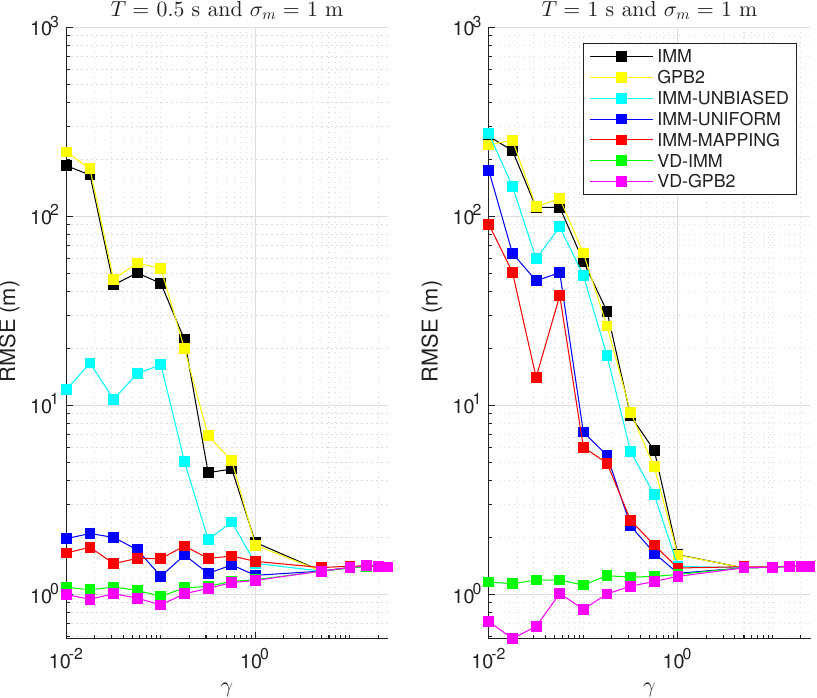}\caption{$\mathrm{RMSE}$ for $\sigma_{m}=1$ m and $T=0.5$ s (left) and
with $T=1$ s (right) for Scenario 1. The VD-GPB2 is the best performing
filter followed by the VD-IMM filter.}\label{fig:RMSE_1m}
\end{figure}

Figure \ref{fig:RMSE_1m} presents the RMSE results with a measurement
noise of $\sigma_{m}=1$ m instead of $\sigma_{m}=0.5$ m (as considered
previously). We can draw the following conclusions:
\begin{enumerate}
\item In general, the RMSE increases with respect to Figure \ref{fig:RMSE_0.5m}
because this simulation uses a measurement noise of $\sigma_{m}=$1
m, whereas the previous simulation used $\sigma_{m}=0.5$ m. For the
proposed methods, this results in an increase of approximately the
same order of magnitude as the increment in the measurement noise.
\item In terms of process noise and update interval, the results remain
mostly similar.
\item The proposed VD-IMM and VD-GPB2 filters also exhibit the best performance
compared with the other approaches.
\end{enumerate}
Across all conditions, as shown in Figures \ref{fig:RMSE_0.5m} and
\ref{fig:RMSE_1m}, the proposed VD-IMM and VD-GPB2 consistently achieve
lower RMSE than all IMM variants and the standard GPB2, demonstrating
superior accuracy and robustness over the entire range of $\gamma$,
measurement noise $\sigma_{m}$ and, sampling rate $T$. Compared
with the unbiased mixing of \cite{Yuan12}, the systematic augmentation
of\,\cite{Granstrom15b}, and the mapping\nobreakdash-based interaction
of \cite{Zubaca22}, the proposed methods achieve improved performance
without requiring tuning or augmentation strategies.

Table \ref{tab:Elapsed-time-simulations} shows the running times
of our Matlab implementations of the filters for the Scenario 1, obtained
on a computer equipped with a 2 GHz Intel(R) Core(TM) Ultra 7 255H
processor. The classical IMM, IMM-unbiased and IMM\nobreakdash-uniform
filters achieve the lowest computational cost, reflecting their reduced
mixing complexity. IMM\nobreakdash-mapping and GPB2 introduce a moderate
overhead due to the additional matrix operations required during the
mixing step and the two\nobreakdash-stage update structure, respectively.
The proposed VD-IMM and VD-GPB2 require higher computational times
due to the variable dimension mixing mechanism.

\begin{table}
\centering
\caption{Running times of the algorithms in ms for Scenario 1.}\label{tab:Elapsed-time-simulations}
\begin{tabular}{lc}
\hline 
Filter &
Running Time {[}ms{]}\tabularnewline
\hline 
IMM &
7.1\tabularnewline
GPB2 &
12.5\tabularnewline
IMM-unbiased &
6.0\tabularnewline
IMM-uniform &
6.6\tabularnewline
IMM-mapping &
8.6 \tabularnewline
VD-IMM &
12.7\tabularnewline
VD-GPB2 &
22.6\tabularnewline
\hline 
\end{tabular}
\end{table}

\subsection{Scenario 2}\label{subsec:Scenario-2}

This Subsection evaluates the performance of the proposed EKF VD-IMM
and EKF VD-GPB2 filters (Section \ref{sec:Extension-to-Nonlinear})
against an EKF implementation of the IMM filter with state augmentation
for heterogeneous motion models presented in \cite{Na2022}. Subsections
\ref{subsec:Dynamic-Model_scen2}, \ref{subsec:Mode-Dynamic-Model_scen2-1}
and \ref{subsec:Measurement-Model_scen2} describe the selected dynamic
model, mode model and measurement model. In Subsections \ref{subsec:Parameters_scen2}
and \ref{subsec:Results_scen2} we present all the parameter configurations
and the results achieved by each filter, respectively.

\subsubsection{Dynamic Model}\label{subsec:Dynamic-Model_scen2}

We consider a 2D CV model in polar coordinates (CV-PC) and a 2D CT
model in Cartesian coordinates (CT-CC) model pair with unequal state
variables and dimensions \cite{Na2022}. The state vector for the
CV-PC model is $x^{\texttt{{CV-PC}}}=[x,y,\theta,v]^{T}$, whereas
for the CT-CC model it is $x^{\texttt{{CT-CC}}}=[x,\dot{x},y,\dot{y},\omega]^{T}$,
where $x,y$ are the position in Cartesian coordinates, $\dot{x},\dot{y}$
are the velocity in Cartesian coordinates, $v$ is the absolute value
of the velocity, $\theta$ is the line of bearing with respect to
the $x$ axis and $\omega$ is the angular velocity. We use three
different modes for these simulations, CV-PC, CT\textsubscript{\selectlanguage{english}%
R}-CC and CT\textsubscript{\selectlanguage{english}%
L}-CC. The nonlinear transition functions, see (\ref{eq:transition_model_non_linear}),
are \cite{Na2022} 
\begin{align}
f^{(\texttt{{CV,CV}})}\left(x\right) & =\begin{bmatrix}x+vT\cos\left(\theta\right)\\
y+vT\sin\left(\theta\right)\\
\theta\\
v
\end{bmatrix}+u^{(\texttt{{CV,CV}})}\\
f^{(\texttt{{CT,CT}})}\left(x\right) & =F^{(\texttt{{CT,CT}})}\left(x\right)x^{(\texttt{{CT,CT}})}+u^{(\texttt{{CT,CT}})}\\
f^{(\texttt{{CV,CT}})}\left(x\right) & =\left[\begin{array}{c}
x+\frac{\dot{x}}{\omega}\sin\left(\omega T\right)-\frac{\dot{y}}{\omega}\left(1-\cos\left(\omega T\right)\right)\\
y+\frac{\dot{y}}{\omega}\sin\left(\omega T\right)+\frac{\dot{x}}{\omega}\left(1-\cos\left(\omega T\right)\right)\\
\arctan\left(\frac{\dot{y}}{\dot{x}}\right)\\
\sqrt{\dot{x}^{2}+\dot{y}^{2}}
\end{array}\right]\nonumber \\
 & \quad+u^{(\texttt{{CV,CT}})}\\
f^{(\texttt{{CT,CV}})}\left(x\right) & =\begin{bmatrix}x+vT\cos\left(\theta\right)\\
v\cos\left(\theta\right)\\
y+vT\sin\left(\theta\right)\\
v\sin\left(\theta\right)\\
0
\end{bmatrix}+u^{(\texttt{{CT,CV}})}
\end{align}
where $T$ is the sampling time and $F^{(\texttt{{CT,CT}})}\left(\cdot\right)$
is given by (\ref{eq:F_CT_CT}). Note that the superscript CV defines
CV-PC and CT defines both CT\textsubscript{\selectlanguage{english}%
R}-CC and CT\textsubscript{\selectlanguage{english}%
L}-CC modes since they are similar except for the offset terms, which
are given by
\begin{align}
u^{(\texttt{{CV,CV}})} & =0_{4\times1}\\
u^{(\texttt{{CT,CT}})} & =0_{5\times1}.\\
u^{(\texttt{{CT\_R,CV}})} & =\left[\begin{array}{c}
0_{4\times1}\\
-\omega^{*}
\end{array}\right]\\
u^{(\texttt{{CT\_L,CV}})} & =\left[\begin{array}{c}
0_{4\times1}\\
\omega^{*}
\end{array}\right]\\
u^{(\texttt{{CT\_R,CT\_L}})} & =\left[\begin{array}{c}
0_{4\times1}\\
-\omega^{*}
\end{array}\right]\\
u^{(\texttt{{CT\_L,CT\_R}})} & =\left[\begin{array}{c}
0_{4\times1}\\
\omega^{*}
\end{array}\right]
\end{align}
where $\omega^{*}$ is a known parameter, as in Scenario 1.

It should be noted that, when there is a change of mode, the choice
of the nonlinear function takes into account kinematic and geometric
considerations such that the change of state is consistent.

The process noise covariance matrices are \cite{Na2022}
\begin{align}
Q^{(\texttt{{CV,CV}})} & =Q^{(\texttt{{CV,CT}})}=G^{(\texttt{{CV}})}\Sigma^{(\texttt{{CV}})}G^{(\texttt{{CV}})T}\\
Q^{(\texttt{{CT,CT}})} & =Q^{(\texttt{{CT,CV}})}=G^{(\texttt{{CT}})}\Sigma^{(\texttt{{CT}})}G^{(\texttt{{CT}})T}
\end{align}

where
\begin{align}
G^{(\texttt{{CV}})} & =\left[\begin{array}{cc}
\frac{T^{2}}{2}\cos\left(\theta\right) & 0\\
\frac{T^{2}}{2}\sin\left(\theta\right) & 0\\
0 & \frac{T^{2}}{2}\\
T & 0
\end{array}\right]\\
G^{(\texttt{{CT}})} & =\left[\begin{array}{ccc}
\frac{T^{2}}{2} & 0 & 0\\
T & 0 & 0\\
0 & \frac{T^{2}}{2} & 0\\
0 & T & 0\\
0 & 0 & T
\end{array}\right]\\
\Sigma^{(\texttt{{CV}})} & \mathrm{=diag}\left(\sigma_{\texttt{{CV}}}^{2},\sigma_{\texttt{{CT}}}^{2}\right)\\
\Sigma^{(\texttt{{CT}})} & \mathrm{=diag}\left(\sigma_{\texttt{{CV}}}^{2},\sigma_{\texttt{{CV}}}^{2},\sigma_{\texttt{{CT}}}^{2}\right)
\end{align}
where $\sigma_{\texttt{{CV}}}^{2}$ and $\sigma_{\texttt{{CT}}}^{2}$
are parameters of the nearly constant velocity model and coordinated
turn model.

At time step $0$, the mode, mean and covariance of the initial Gaussian
distribution, see (\ref{eq:time_0_gaussian}), are
\begin{align}
f_{0|0}(\texttt{{CV}}) & =1\\
\overline{x}_{0|0}^{(\texttt{{CV}})} & =\left[\begin{array}{c}
0\left(\mathrm{m}\right),0\left(\mathrm{m}\right),15\times\frac{\pi}{180}\left(\mathrm{rad}\right),50\left(\frac{\mathrm{m}}{\mathrm{s}}\right)\end{array}\right]^{T}\\
P_{0|0}^{(\texttt{{CV}})} & =\mathrm{diag\left(\left[1\left(\mathrm{m^{2}}\right),1\left(\mathrm{m^{2}}\right),0.01\left(\mathrm{rad}^{2}\right),0.1\left(\mathrm{m^{2}}/\mathrm{s^{2}}\right)\right]\right).}
\end{align}

\subsubsection{Mode Dynamic Model}\label{subsec:Mode-Dynamic-Model_scen2-1}

In this scenario, we employ the same dynamic model presented in Subsection
\ref{subsec:Sim_Mode-Model}.

\subsubsection{Measurement Model}\label{subsec:Measurement-Model_scen2}

The target position is observed through a noisy measurement $z\in\mathbb{R}^{2}$,
leading to the following measurement matrices
\begin{align}
H^{(\texttt{{CV-PC}})} & =\left[\begin{array}{cccc}
1 & 0 & 0 & 0\\
0 & 1 & 0 & 0
\end{array}\right]\\
H^{(\texttt{{CT-CC}})} & =\left[\begin{array}{ccccc}
1 & 0 & 0 & 0 & 0\\
0 & 0 & 1 & 0 & 0
\end{array}\right]
\end{align}
and $d^{\left(r_{k}\right)}=0_{2,1}$. The measurement covariance
matrix $R_{k}$ is 
\begin{equation}
R^{\left(r_{k}\right)}=\sigma_{m}^{2}I_{2}.
\end{equation}

\subsubsection{Parameters}\label{subsec:Parameters_scen2}

We set the following parameters: $T=1$ s, $\omega^{*}=15\times\frac{\pi}{180}$
rad/s, $\sigma'_{\texttt{{CV}}}=1$ m/s\textsuperscript{2}, $\sigma'_{\texttt{{CT}}}=1\times\frac{\pi}{180}$
rad\textsuperscript{}/s\textsuperscript{1/2} $\gamma=10^{-2}$,
and $\sigma_{m}=0.5$ m.

\subsubsection{Results}\label{subsec:Results_scen2}

\begin{figure}
\includegraphics[scale=0.6]{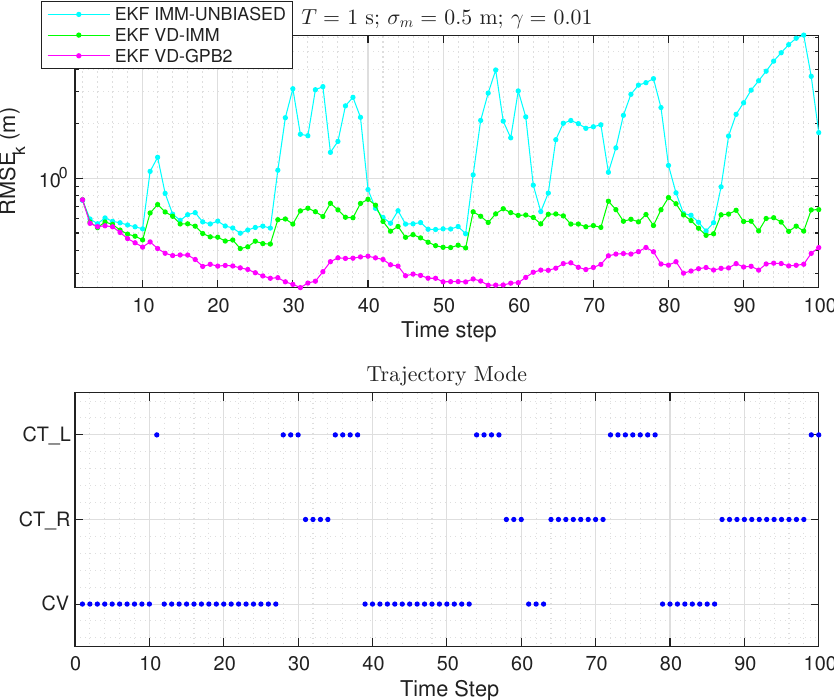}\caption{RMSE for $T=1$ s, $\sigma_{m}=0.5$ m and $\gamma=0.01$ and trajectory
mode ground truth for Scenario 2. The EKF VD-GPB2 is the best performing
filter followed by the EKF VD-IMM filter.}\label{fig:RMSE_secenario_2}
\end{figure}

In Figure\,\ref{fig:RMSE_secenario_2}, we present the RMSE against
time for the considered algorithms. From the beginning of the simulation,
the proposed filters outperform the augmented EKF IMM-unbiased filter.
As expected, the best performing filter is the EKF VD-GPB2 filter,
followed by the EKF VD-IMM filter. The EKF VD-IMM and EKF IMM-unbiased
filters show a higher $\mathrm{{RMSE}}$ than the EKF VD-GPB2 filter
during transitions between dynamic modes. Nevertheless, EKF VD-IMM
filter always has a lower error than the EKF IMM unbiased filter.
Therefore, Figure \ref{fig:RMSE_secenario_2} shows that the proposed
VD filters can outperform a state-of-the-art IMM filter specifically
designed for heterogeneous motion models.

Table \ref{tab:Elapsed-time-simulations-Scenario_2} shows the running
times of our Matlab implementations of the filters for the Scenario
2. As in Scenario 1, the proposed EKF VD-IMM and EKF VD-GPB2 filters
require higher computational times compared to the EKF IMM-unbiased
filter due to the variable dimension mixing mechanism.

\begin{table}
\centering
\caption{Running times of the algorithms in ms for Scenario 2.}\label{tab:Elapsed-time-simulations-Scenario_2}
\begin{tabular}{lc}
\hline 
Filter &
Running Time {[}ms{]}\tabularnewline
EKF IMM-unbiased &
9.8\tabularnewline
EKF VD-IMM &
19.5 \tabularnewline
EKF VD-GPB2 &
26.4\tabularnewline
\hline 
\end{tabular}
\end{table}

\section{Conclusions}\label{sec:Conclusions}

In this paper, we have proposed a mathematically principled solution
to the problem of multiple model filtering with states of variable
dimensionality. We have developed a principled Bayesian filtering
formulation with states of variable dimensionality as well as principled
derivations of IMM and GPB2 filters with variable dimensionality based
on obtaining a Gaussian density for each mode via the minimisation
of the Kullback-Leibler divergence.

The resulting filters are shown to outperform standard IMM algorithms
that were adapted to handle states of different dimensionality, especially
in situations with low process noise at the expense of a higher computational
burden.

Future work can extend the proposed variable dimension filters to
sigma-point based implementations, multi-sensor fusion and multi-target
tracking.

\bibliographystyle{IEEEtran}
\bibliography{9C__Users_rober_Documents_Doctorado_Bibtex_references}

\cleardoublepage{}

{\LARGE Supplemental material:} {\LARGE ``Variable Dimension IMM and
GPB2 Filters for Tracking in Multiple Model Systems''}{\LARGE\par}

\appendices{}

\section{}\label{sec:appendix_a}

This appendix proves the Bayesian filtering prediction and update
steps presented in Lemmas \ref{lem:bayesian_pred} and \ref{lem:bayesian_upd}.
To do so, we first review the concept of integration and density on
the space $\mathbb{X}$.

It should be noted that the resulting variable dimension prediction
and update steps turn out to be similar to the prediction and update
steps of a fixed dimensional system, with the difference that densities,
including the transition density (which must depend on $r_{k}$ and
$r_{k-1}$ for the variable dimensional system) and the measurement
density, are defined for variable dimensional states. Nevertheless,
for completeness, it is important to show this result using the variable
dimensional space with its integral, its corresponding Chapman-Kolmogorov
equation and Bayes' update.

\subsection{Integrals on the Variable Dimension Space}

Given a real-valued function $\pi\left(\cdot\right)$ on space $\mathbb{X}$,
its integral (from a finite-set statistics point of view) is given
by \cite[Eq. (3.50)]{Mahler2014}
\begin{align}
\int_{\mathbb{X}}\pi\left(\widetilde{x}\right)d\widetilde{x} & =\sum_{r=1}^{m}\int_{\mathbb{R}^{n_{r}}}\pi\left(r,x\right)dx\label{eq:joint_integral}
\end{align}
where $\widetilde{x}=\left(r,x\right)\in\mathbb{X}$.

A function $\pi\left(\cdot\right)$ is a density on space $\mathbb{X}$
if $\pi\left(\cdot\right)\geq0$ and its integral is one. In this
case, we can also write
\begin{align}
\int_{\mathbb{X}}\pi\left(\widetilde{x}\right)d\widetilde{x} & =\sum_{r=1}^{m}\pi\left(r\right)\int_{\mathbb{R}^{n_{r}}}\pi\left(x|r\right)dx\label{eq:marginal_integral}
\end{align}
where $\pi\left(r\right)$ is the marginal density of $r$ and $\pi\left(x|r\right)$
is the conditional distribution of $x$ given $r$.

Note that a density on space $\mathbb{X}$ characterises a random
variable on space $\mathbb{X}$. It should be noted that a measure-theoretic
integral can also be defined by using a unitless Lebesgue measure
for each sub-space, as was done in Equations (A.5)-(A.6) in \cite{Xia2019}.
Nevertheless, both the finite-set statistic and measure-theoretic
integrals lead to the same results, and we use (\ref{eq:joint_integral})
for notational simplicity. We also drop the space of integration from
the integrals since it is clear from context.

\subsection{Prediction }\label{subsec:Prediction}

The prediction step in Lemma \ref{lem:bayesian_pred} is obtained
by applying the Chapman-Kolmogorov equation on space $\mathbb{X}$,
using the integral (\ref{eq:joint_integral}). That is, the predicted
density is \cite{Mahler2014}
\begin{align}
f_{k|k-1}\left(\widetilde{x}_{k}\right) & =\int p\left(\widetilde{x}_{k}|\widetilde{x}_{k-1}\right)f_{k-1|k-1}\left(\widetilde{x}_{k-1}\right)d\widetilde{x}_{k-1}.\label{eq:prediction_integral}
\end{align}

Then, substituting Equation (\ref{eq:joint_mode_state}) in (\ref{eq:prediction_integral})
and applying the integral (\ref{eq:marginal_integral}) yields
\begin{align}
f_{k|k-1}\left(r_{k},x_{k}\right) & =\sum_{r_{k-1}=1}^{m}\int p\left(r_{k},x_{k}|r_{k-1},x_{k-1}\right)\nonumber \\
 & \quad\times f_{k-1|k-1}\left(r_{k-1},x_{k-1}\right)dx_{k-1}\nonumber \\
 & =\sum_{r_{k-1}=1}^{m}\mu\left(r_{k}|r_{k-1}\right)\int\pi\left(x_{k}|x_{k-1},r_{k},r_{k-1}\right)\nonumber \\
 & \quad\times f_{k-1|k-1}\left(r_{k-1},x_{k-1}\right)dx_{k-1}\nonumber \\
 & =\sum_{r_{k-1}=1}^{m}\mu\left(r_{k}|r_{k-1}\right)f_{k-1|k-1}\left(r_{k-1}\right)\nonumber \\
 & \quad\times\int\pi\left(x_{k}|x_{k-1},r_{k},r_{k-1}\right)\nonumber \\
 & \quad\times f_{k-1|k-1}\left(x_{k-1}|r_{k-1}\right)dx_{k-1}.\label{eq:joint_prediction}
\end{align}
As shown in (\ref{eq:joint_prediction}), in the multiple mode approach,
the contribution of each mode $r_{k-1}$ is incorporated through the
summation operator. By marginalizing the previous equation over $x_{k}$,
the predicted density of the mode can be obtained as follows
\begin{align}
f_{k|k-1}\left(r_{k}\right) & =\int f_{k|k-1}\left(r_{k},x_{k}\right)dx_{k}\nonumber \\
 & =\sum_{r_{k-1}=1}^{m}\mu\left(r_{k}|r_{k-1}\right)f_{k-1|k-1}\left(r_{k-1}\right).
\end{align}
Finally, the predicted density of the state given the mode can be
computed using the conditional probability theorem applied to equations
(\ref{eq:joint_prediction}) and (\ref{eq:mode_prediction}), yielding
\begin{align}
f_{k|k-1}\left(x_{k}|r_{k}\right) & =\frac{f_{k|k-1}\left(r_{k},x_{k}\right)}{f_{k|k-1}\left(r_{k}\right)}\nonumber \\
 & =\frac{1}{\sum_{r_{k-1}=1}^{m}\mu\left(r_{k}|r_{k-1}\right)f_{k-1|k-1}\left(r_{k-1}\right)}\nonumber \\
 & \quad\times\sum_{r_{k-1}=1}^{m}\mu\left(r_{k}|r_{k-1}\right)f_{k-1|k-1}\left(r_{k-1}\right)\nonumber \\
 & \quad\times\int\pi\left(x_{k}|x_{k-1},r_{k},r_{k-1}\right)\nonumber \\
 & \quad\times f_{k-1|k-1}\left(x_{k-1}|r_{k-1}\right)dx_{k-1}.\label{eq:state_prediction_prev}
\end{align}

Then, equation (\ref{eq:state_prediction_prev}) can be rewritten
as
\begin{align}
f_{k|k-1}\left(x_{k}|r_{k}\right) & =\sum_{r_{k-1}=1}^{m}\alpha_{k|k-1}^{\left(r_{k},r_{k-1}\right)}\int\pi\left(x_{k}|x_{k-1},r_{k},r_{k-1}\right)\nonumber \\
 & \quad\times f_{k-1|k-1}\left(x_{k-1}|r_{k-1}\right)dx_{k-1}
\end{align}
where $\alpha_{k|k-1}^{\left(r_{k},r_{k-1}\right)}$ is given by (\ref{eq:IMM_alpha}).
This finishes the proof of Lemma \ref{lem:bayesian_pred}.

\subsection{Update }\label{subsec:Update}

Lemma \ref{lem:bayesian_upd} is proved using Bayes' rule applied
to the predicted density on space $\mathbb{X}$. Accordingly, the
posterior density of the mode $r_{k}$ and the state $x_{k}$ at time
step $k$ is given by \cite{Mahler2014}
\begin{align}
 & f_{k|k}\left(r_{k},x_{k}\right)\nonumber \\
 & =\frac{l\left(z_{k}|r_{k},x_{k}\right)f_{k|k-1}\left(r_{k},x_{k}\right)}{\sum_{r_{k}=1}^{m}\int l\left(z_{k}|r_{k},x_{k}\right)f_{k|k-1}\left(r_{k},x_{k}\right)dx_{k}}\nonumber \\
 & =\frac{f_{k|k-1}\left(r_{k}\right)l\left(z_{k}|r_{k},x_{k}\right)f_{k|k-1}\left(x_{k}|r_{k}\right)}{\sum_{r_{k}=1}^{m}f_{k|k-1}\left(r_{k}\right)\int l\left(z_{k}|r_{k},x_{k}\right)f_{k|k-1}\left(x_{k}|r_{k}\right)dx_{k}}.\label{eq:joint_update}
\end{align}

In the above equation, the first term in the numerator, $l\left(z_{k}|r_{k},x_{k}\right)$,
represents the likelihood of the measurement $z_{k}$ given the mode
$r_{k}$ and the state $x_{k}$ at time step $k$. The second term,
$f_{k|k-1}\left(r_{k},x_{k}\right)$, corresponds to the predicted
joint probability of the mode $r_{k}$ and the state $x_{k}$ at time
step $k$, conditioned on time step $k-1$. The denominator acts as
a normalization constant, ensuring that the resulting expression is
a valid density.

By marginalizing equation (\ref{eq:joint_update}) over the $x_{k}$
dimension, the posterior density of the mode is obtained as follows
\begin{align}
 & f_{k|k}\left(r_{k}\right)\nonumber \\
 & =\int f_{k|k}\left(r_{k},x_{k}\right)dx_{k}\nonumber \\
 & =\frac{f_{k|k-1}\left(r_{k}\right)\int l\left(z_{k}|r_{k},x_{k}\right)f_{k|k-1}\left(x_{k}|r_{k}\right)dx_{k}}{\sum_{r_{k}=1}^{m}f_{k|k-1}\left(r_{k}\right)\int l\left(z_{k}|r_{k},x_{k}\right)f_{k|k-1}\left(x_{k}|r_{k}\right)dx_{k}}.
\end{align}
Finally, the posterior density of the state given the mode is obtained
by applying the conditional probability theorem to equations (\ref{eq:joint_update})
and (\ref{eq:update_mode}), which leads to the following expression
\begin{align}
f_{k|k}\left(x_{k}|r_{k}\right) & =\frac{f_{k|k}\left(x_{k},r_{k}\right)}{f_{k|k}\left(r_{k}\right)}\nonumber \\
 & =\frac{l\left(z_{k}|r_{k},x_{k}\right)f_{k|k-1}\left(x_{k}|r_{k}\right)}{\int l\left(z_{k}|r_{k},x_{k}\right)f_{k|k-1}\left(x_{k}|r_{k}\right)dx_{k}}.
\end{align}

This finishes the proof of Lemma \ref{lem:bayesian_upd}.

\section{}\label{sec:appendix_b}

This appendix provides the expression of the KLD on space $\mathbb{X}$
and also provides a KLD minimisation result that is used in the proofs
of the VD-IMM prediction step and the VD-GPB2 update step.

\subsection{KLD on space $\mathbb{X}$}

Making use of the integral on space $\mathbb{X}$, given by (\ref{eq:marginal_integral}),
the KLD between densities $\pi(\cdot)$ and $q(\cdot)$ on space $\mathbb{X}$
can be written as
\begin{align}
D\left(\pi||q\right) & =\int\pi\left(\widetilde{x}\right)\log\frac{\pi\left(\widetilde{x}\right)}{q\left(\widetilde{x}\right)}d\widetilde{x}\nonumber \\
 & =\sum_{r=1}^{m}\pi\left(r\right)\int\pi\left(x|r\right)\log\frac{\pi\left(r\right)\pi\left(x|r\right)}{q\left(r\right)q\left(x|r\right)}dx\nonumber \\
 & =\sum_{r=1}^{m}\pi\left(r\right)\log\frac{\pi\left(r\right)}{q\left(r\right)}\nonumber \\
 & \quad+\sum_{r=1}^{m}\pi\left(r\right)\int\pi\left(x|r\right)\log\frac{\pi\left(x|r\right)}{q\left(x|r\right)}dx.\label{eq:KLD_pi_q}
\end{align}

Note that this is the KLD between the probability mass functions $\pi\left(r\right)$
and $q\left(r\right)$ plus the weighted KLD between the densities
$\pi\left(x|r\right)$ and $q\left(x|r\right)$.

\subsection{KLD Minimisation on Space $\mathbb{X}$}

In this appendix, we show the following lemma regarding KLD minimisation
on space $\mathbb{X}$.
\begin{lem}
\label{lem:KLD_minimisation}Given a density $\pi\left(\cdot\right)$
on space $\mathbb{X}$ such that the density of the state given the
mode $r$ is a Gaussian mixture of the form
\begin{align}
\pi\left(x|r\right) & =\sum_{j=1}^{J^{(r)}}w_{j}^{\left(r\right)}\mathcal{N}\left(x;\overline{x}_{j}^{\left(r\right)},P_{j}^{\left(r\right)}\right)\label{eq:pi_x_r_append}
\end{align}
where $J^{(r)}$ is the number of components and $w_{j}^{\left(r\right)}$,
$\overline{x}_{j}^{\left(r\right)}$ and $P_{j}^{\left(r\right)}$
are the weight, mean and covariance matrix of the $j$-th component,
the density $q\left(\cdot\right)$ whose density of the state given
the mode $r$ is Gaussian of the form
\begin{equation}
q\left(x|r\right)=\mathcal{N}\left(x;\overline{x}^{\left(r\right)},P^{\left(r\right)}\right)
\end{equation}
and minimises the KLD $D\left(\pi||q\right)$ in \cite{Bishop2006,Hershey2007}
is characterised by
\begin{align}
q\left(r\right) & =\pi\left(r\right)\\
\overline{x}^{\left(r\right)} & =\sum_{j=1}^{J^{(r)}}w_{j}^{\left(r\right)}\overline{x}_{j}^{\left(r\right)}\label{eq:mean_moment_matching_append}\\
P^{\left(r\right)} & =\sum_{j=1}^{J^{(r)}}w_{j}^{\left(r\right)}\left(\overline{x}_{j}^{\left(r\right)}-\overline{x}^{\left(r\right)}\right)\left(\overline{x}_{j}^{\left(r\right)}-\overline{x}^{\left(r\right)}\right)^{T}\nonumber \\
 & \quad+\sum_{j=1}^{J^{(r)}}w_{j}^{\left(r\right)}P_{j}^{\left(r\right)}.\label{eq:cov_moment_matching_append}
\end{align}
\end{lem}
That is, to minimise the KLD, we have the same distribution for the
mode, and we perform moment matching for each mode. The proof of this
lemma is as follows. 

From the first term in Equation (\ref{eq:KLD_pi_q}), it is evident
that the KLD is minimised by setting $q\left(r\right)=\pi\left(r\right)$.
Now we proceed to minimise the KLD over $\overline{x}^{\left(r\right)}$
and $P^{\left(r\right)}$ for all $r$. These can be obtained by solving
\begin{align}
\operatorname*{argmin}_{\overline{x}^{\left(1\right)},P^{\left(1\right)},...,\overline{x}^{\left(m\right)},P^{\left(m\right)}} & -\sum_{r=1}^{m}\pi\left(r\right)\nonumber \\
 & \times\int\pi\left(x|r\right)\log\mathcal{N}\left(x;\overline{x}^{\left(r\right)},P^{\left(r\right)}\right)dx.
\end{align}

As we have to minimise the mean and covariance for each $r$ and the
function is additive over these components, we can just solve
\begin{equation}
\operatorname*{argmin}_{\overline{x}^{\left(r\right)},P^{\left(r\right)}}-\int\pi\left(x|r\right)\log\mathcal{N}\left(x;\overline{x}^{\left(r\right)},P^{\left(r\right)}\right)dx.
\end{equation}

As $\pi\left(x|r\right)$ is of the form (\ref{eq:pi_x_r_append}),
by standard KLD minimisation for Gaussian mixtures \cite{Bishop2006,Hershey2007},
we obtain that $\overline{x}^{\left(r\right)}$ and $P^{\left(r\right)}$
are given by moment matching, which corresponds to (\ref{eq:mean_moment_matching_append})
and (\ref{eq:cov_moment_matching_append}). This finishes the proof
of Lemma \ref{lem:KLD_minimisation}.

\section{}\label{sec:appendix_c}

This appendix proves the VD-IMM filter prediction and update steps,
presented in Theorem \ref{thm:prediction} and Theorem \ref{thm:update}.

\subsection{Prediction}\label{subsec:VDIMM_prediction_proof}

We consider that each mode has its own Gaussian approximation such
that the posterior at time $k-1$ is
\begin{align}
 & f_{k-1|k-1}\left(x_{k-1}|r_{k-1}\right)\nonumber \\
 & =\mathcal{N}\left(x_{k-1};\overline{x}_{k-1|k-1}^{\left(r_{k-1}\right)},P_{k-1|k-1}^{\left(r_{k-1}\right)}\right).\label{eq:state_posterior_k_1}
\end{align}

The predicted density for the mode is then given by (\ref{eq:mode_prediction}).
The predicted density for the state given the mode is given by (\ref{eq:state_prediction}).
By standard Kalman filter prediction \cite{Sarkka_book23}, we first
have that
\begin{align}
 & \int\pi\left(x_{k}|x_{k-1},r_{k},r_{k-1}\right)f_{k-1|k-1}\left(x_{k-1}|r_{k-1}\right)dx_{k-1}\nonumber \\
 & =\mathcal{N}\left(x_{k};\overline{x}_{k|k-1}^{\left(r_{k},r_{k-1}\right)},P_{k|k-1}^{\left(r_{k},r_{k-1}\right)}\right)
\end{align}
where

\begin{align}
\overline{x}_{k|k-1}^{\left(r_{k},r_{k-1}\right)} & =F^{\left(r_{k},r_{k-1}\right)}\overline{x}_{k-1|k-1}^{\left(r_{k-1}\right)}+b^{(r_{k},r_{k-1})}\label{eq:IMM_state_prediction}\\
P_{k|k-1}^{\left(r_{k},r_{k-1}\right)} & =F^{\left(r_{k},r_{k-1}\right)}P_{k-1|k-1}^{\left(r_{k-1}\right)}\left(F^{\left(r_{k},r_{k-1}\right)}\right)^{T}\nonumber \\
 & \quad+Q^{\left(r_{k},r_{k-1}\right)}.\label{eq:IMM_covariance_prediction}
\end{align}
This results in
\begin{align}
 & f_{k|k-1}\left(x_{k}|r_{k}\right)\nonumber \\
 & =\sum_{r_{k-1}=1}^{m}\alpha_{k|k-1}^{\left(r_{k},r_{k-1}\right)}\mathcal{N}\left(x_{k};\overline{x}_{k|k-1}^{\left(r_{k},r_{k-1}\right)},P_{k|k-1}^{\left(r_{k},r_{k-1}\right)}\right).\label{eq:predicted_gaussian_mixture}
\end{align}

Applying Lemma \ref{lem:KLD_minimisation}, we can approximate (\ref{eq:predicted_gaussian_mixture})
as a Gaussian distribution by KLD minimisation, using (\ref{eq:mean_moment_matching_append})
and (\ref{eq:cov_moment_matching_append}). This results in (\ref{eq:IMM_x_pred_KLD})-(\ref{eq:IMM_P_pred_KLD})
providing the proof of Theorem \ref{thm:prediction}.

\subsection{Update}

We consider that the predicted density of the state given each mode
is Gaussian, given by (\ref{eq:predicted_updated_density}). The posterior
density of the state given the mode is obtained by plugging (\ref{eq:predicted_updated_density})
and (\ref{eq:measurement_model}) into the update equation (\ref{eq:update_state}).
This yields
\begin{align}
f_{k|k}\left(x_{k}|r_{k}\right) & \propto\mathcal{N}\left(z_{k};H^{\left(r_{k}\right)}x_{k}+d^{(r_{k})},R^{\left(r_{k}\right)}\right)\nonumber \\
 & \times\mathcal{N}\left(x_{k};\overline{x}_{k|k-1}^{\left(r_{k}\right)},P_{k|k-1}^{\left(r_{k}\right)}\right).
\end{align}

Using the Kalman filter update \cite{Sarkka_book23} yields
\begin{align}
f_{k|k}\left(x_{k}|r_{k}\right) & =\mathcal{N}\left(x_{k};\overline{x}_{k|k}^{\left(r_{k}\right)},P_{k|k}^{\left(r_{k}\right)}\right)\label{eq:update_state_Gaussian}
\end{align}
where
\begin{align}
\overline{x}_{k|k}^{\left(r_{k}\right)} & =\overline{x}_{k|k-1}^{\left(r_{k}\right)}+P_{k|k-1}^{\left(r_{k}\right)}\left(H^{\left(r_{k}\right)}\right)^{T}\left(S^{\left(r_{k}\right)}\right)^{-1}\nonumber \\
 & \quad\times\left(z_{k}-\hat{z}^{\left(r_{k}\right)}\right)\\
P_{k|k}^{\left(r_{k}\right)} & =P_{k|k-1}^{\left(r_{k}\right)}-P_{k|k-1}^{\left(r_{k}\right)}\left(H^{\left(r_{k}\right)}\right)^{T}\left(S^{\left(r_{k}\right)}\right)^{-1}\nonumber \\
 & \quad\times H^{\left(r_{k}\right)}P_{k|k-1}^{\left(r_{k}\right)}\\
\hat{z}^{\left(r_{k}\right)} & =H^{\left(r_{k}\right)}\overline{x}_{k|k-1}^{\left(r_{k}\right)}+d^{(r_{k})}\\
S^{\left(r_{k}\right)} & =H^{\left(r_{k}\right)}P_{k|k-1}^{\left(r_{k}\right)}\left(H^{\left(r_{k}\right)}\right)^{T}+R^{\left(r_{k}\right)}.
\end{align}

The posterior of the mode is obtained plugging (\ref{eq:measurement_model})
into (\ref{eq:update_mode}), which yields
\begin{align}
f_{k|k}\left(r_{k}\right) & =\frac{f_{k|k-1}\left(r_{k}\right)\mathcal{N}\left(z_{k};\hat{z}^{\left(r_{k}\right)},S^{\left(r_{k}\right)}\right)}{\sum_{r_{k}=1}^{m}f_{k|k-1}\left(r_{k}\right)\mathcal{N}\left(z_{k};\hat{z}^{\left(r_{k}\right)},S^{\left(r_{k}\right)}\right)}.
\end{align}
This finishes the proof of Theorem \ref{thm:update}.

\section{}\label{sec:appendix_d}

This appendix proves the variable dimension VD-GPB2 filtering equations
presented in Theorem \ref{thm:prediction_gpb2} and Theorem \ref{thm:update_gpb2}.

\subsection{Prediction}

We consider that the posterior of each mode at time $k-1$ is Gaussian
of the form (\ref{eq:state_posterior_k_1}). The predicted density
for the mode is given by (\ref{eq:mode_prediction}). The predicted
density for the state given the mode is given by plugging (\ref{eq:transition_model})
into (\ref{eq:state_prediction}). As was proved in Section \ref{subsec:VDIMM_prediction_proof},
this results in a predicted density of the state $x_{k}$ given the
mode $r_{k}$ given by (\ref{eq:predicted_gaussian_mixture}). This
finishes the proof of Theorem \ref{thm:prediction_gpb2}.

\subsection{Update}

We consider that the predicted density of the state given each mode
is a Gaussian mixture, given by (\ref{eq:predicted_gaussian_mixture}).
The posterior density of the state given the mode is obtained by plugging
(\ref{eq:predicted_gaussian_mixture}) and (\ref{eq:measurement_model})
into the update equation (\ref{eq:update_state}). This yields
\begin{align}
f_{k|k}\left(x_{k}|r_{k}\right) & \propto\sum_{r_{k-1}=1}^{m}\alpha_{k|k-1}^{\left(r_{k},r_{k-1}\right)}\nonumber \\
 & \times\mathcal{N}\left(z_{k};H^{\left(r_{k}\right)}x_{k}+d^{(r_{k})},R^{\left(r_{k}\right)}\right)\nonumber \\
 & \times\mathcal{N}\left(x_{k};\overline{x}_{k|k-1}^{\left(r_{k},r_{k-1}\right)},P_{k|k-1}^{\left(r_{k},r_{k-1}\right)}\right).
\end{align}

We know from the Kalman filter update \cite{Sarkka_book23} that we
can write
\begin{align}
f_{k|k}\left(x_{k}|r_{k}\right) & \propto\sum_{r_{k-1}=1}^{m}\alpha_{k|k-1}^{\left(r_{k},r_{k-1}\right)}\nonumber \\
 & \times\mathcal{N}\left(z_{k};\hat{z}^{\left(r_{k},r_{k-1}\right)},S^{\left(r_{k},r_{k-1}\right)}\right)\nonumber \\
 & \times\mathcal{N}\left(x_{k};\overline{x}_{k|k}^{\left(r_{k},r_{k-1}\right)},P_{k|k}^{\left(r_{k},r_{k-1}\right)}\right)
\end{align}
where $\hat{z}^{\left(r_{k},r_{k-1}\right)}$, $S^{\left(r_{k},r_{k-1}\right)}$,
$\overline{x}_{k|k}^{\left(r_{k},r_{k-1}\right)}$ and $P_{k|k}^{\left(r_{k},r_{k-1}\right)}$
are defined in Theorem \ref{thm:update_gpb2}. This yields
\begin{multline}
f_{k|k}\left(x_{k}|r_{k}\right)\\
=\sum_{r_{k-1}=1}^{m}\rho^{(r_{k},r_{k-1})}\mathcal{N}\left(x_{k};\overline{x}_{k|k}^{\left(r_{k},r_{k-1}\right)},P_{k|k}^{\left(r_{k},r_{k-1}\right)}\right)\label{eq:update_state_density_append}
\end{multline}
where $\rho^{\left(r_{k},r_{k-1}\right)}$ is given by (\ref{eq:rho}).

We can approximate (\ref{eq:update_state_density_append}) as a Gaussian
via moment matching by using (\ref{eq:mean_moment_matching_append})
and (\ref{eq:cov_moment_matching_append}). The corresponding updated
mean and covariance matrix are $\overline{x}_{k|k}^{\left(r_{k}\right)}$
and $P_{k|k}^{\left(r_{k}\right)}$ in (\ref{eq:GPB2_x_upd_KLD})
and (\ref{eq:GPB2_P_upd_KLD}). 

On the other hand, the posterior of the mode $f_{k|k}\left(r_{k}\right)$
is obtained by plugging (\ref{eq:predicted_gaussian_mixture}) and
(\ref{eq:measurement_model}) into (\ref{eq:update_mode}). By using
the properties of the Gaussian densities \cite{Sarkka_book23}, this
results in (\ref{eq:update_mode_gpb2}). This finishes the proof of
Theorem \ref{thm:update_gpb2}.
\end{document}